\documentclass[notoc]{JHEP3}

\usepackage[dvips]{graphicx}
\usepackage{axodraw,epsfig,amssymb,bm,amsmath}

\newcommand{\bc}{\begin{center}}
\newcommand{\ec}{\end{center}}

\title{Vector Meson Photoproduction
from the BFKL Equation II: Phenomenology}

\author{G.G\ Poludniowski$\,^a$ , R.\ Enberg$\,^b$, J.R.\ Forshaw$\,^a$
and L.\ Motyka$\,^c$
\\
$^a$ Department of Physics \& Astronomy, University of Manchester,\\ 
\phantom{$^b$} Manchester M13 9PL, UK
\\
$^b$ High Energy Physics, Uppsala University, Box 535, SE-751 21 Uppsala, 
Sweden 
\\ 
$^c$ Institute of Physics, Jagellonian University, Reymonta 4, 30-059 Krak\'ow, Poland}

\abstract{Diffractive vector meson photoproduction accompanied by proton
dissociation is studied for large momentum transfer.
The process is described by the non-forward BFKL equation which we use
to compare to data collected at the HERA collider. \\
}
\preprint{TSL/ISV-2003-XYZ}

\usepackage{amsmath}
\usepackage{amssymb}
\usepackage{epsfig}

\newcommand{\be}{\begin{equation}}
\newcommand{\ee}{\end{equation}}

\newcommand{\f}{\ensuremath{\,_2F_1}}

\newcommand{\ba}{\begin{eqnarray}}
\newcommand{\ea}{\end{eqnarray}}

\def\repart{\mathcal{R}\mathrm{e}\,}

\def\mq{m}

\newcommand{\beq}[1]{\begin{equation}\label{#1}}
\newcommand{\eeq}{\end{equation}}
\newcommand{\beqar}[1]{\begin{eqnarray}\label{#1}}
\newcommand{\eeqar}{\end{eqnarray}}

\begin{document}

\section{Introduction}

In a previous paper, \cite{EFMP}, we performed a detailed
theoretical study of the diffractive production of 
vector mesons at high momentum transfer in $\gamma p$ collisions. 
We worked throughout in the leading logarithmic BFKL framework \cite{BFKL}
and factorised the meson production from the hard subprocess using a
set of meson light-cone wavefunctions. 

\begin{figure}
{\par\centering {\includegraphics[height=4.5cm,angle=0]{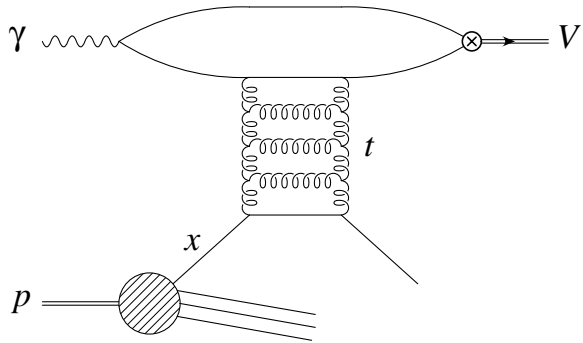}} 
\par}
\caption{Diffractive vector meson photoproduction at large momentum transfer.}
\label{fig1}
\end{figure}

The process is illustrated in Figure \ref{fig1}. The photon and proton 
collide to produce a final state containing a vector meson and the products 
of a proton dissociation. The meson is produced with large transverse 
momentum, the square of which is equal to the Mandelstam variable $-t$, and 
this is in turn much smaller than the available centre-of-mass energy $s$, 
i.e. $s \gg -t \gg \Lambda_{\mathrm{QCD}}^2.$ This hierarchy ensures that the
meson and the proton dissociation are far apart in rapidity.
Note, since we will always integrate over the 
products of the proton dissociation we do not exclude the possibility that 
the proton may scatter elastically. However, this contribution is negligibly 
small for sufficiently large $t$ whence the
finally state is dominated by configurations containing a single jet with
transverse momentum balancing that of the meson \cite{FR}. 

This process has been subject to considerable experimental investigation in
recent years, with the measurement of the $t$-distribution of the meson
($\rho$, $\phi$ and $J/\Psi$)
\cite{ZEUS1,Aktas:2003zi} and the spin-density matrix elements extracted from 
its decay \cite{ZEUS1,Aktas:2003zi} 
being the highlights. It is the purpose of this
paper to compare the theoretical results of \cite{EFMP} to the data.
Before doing so, we recall the status of diffractive meson production
at high-$t$ prior to this analysis.

The experimental data indicate that the cross-section $d\sigma/dt \sim
1/t^3$ for $\rho$ and $\phi$ production. Moreover, the relative smallness
of the $r^{04}_{00}$ spin-density matrix element suggests that the 
interactions tend to produce transversely polarised mesons. This is in stark
contrast with simple theoretical expectations. Lowest order perturbative
QCD does indeed predict a $\sim 1/t^3$ distribution but for longitudinally
polarised mesons. For transversely polarised mesons, the expected
distribution is $\sim 1/t^4$. This observation has led the authors of
\cite{IKSS} to postulate that the
production of transverse mesons is enhanced by the presence of a large
non-perturbative coupling of the photon to chiral odd quark-antiquark
configurations\footnote{We refer to ``chiral even'' and ``chiral odd''
configurations to denote the coupling of the photon to a quark and antiquark
of equal (chiral even) and opposite (chiral odd) chirality.
The chiral odd coupling vanishes for a pointlike coupling to massless
quarks.}. The authors of \cite{IKSS} also identified the 
breakdown of perturbative factorisation in the two-gluon exchange model
due to divergent contributions from configurations where either the quark
or antiquark carry all the momentum of the meson (i.e. end-point 
contributions). They argued that this divergent behaviour would be tamed 
by Sudakov effects and that the dominant chiral odd amplitude was 
in any case free of problems. In contrast, the authors of \cite{Hoyer} 
make a qualitative argument that it is precisely these end-point 
contributions which are responsible for the $1/t^3$ behaviour seen in data 
rather than the $1/t^4$ which would be expected in perturbation theory.
Finally, we should also comment on the apparently naive approach
of \cite{FP}. In this analysis, longitudinal meson production is suppressed
by virtue of the fact that the quark and antiquark share equally the
longitudinal momentum of the meson, and transverse meson production is
enhanced due to the use of the constituent quark mass. Consequently, 
in what follows we
shall take care to comment on each of the above analyses as appropriate
and we shall pay particularly close attention to the end-point behaviour;
there being no end-point divergences in the BFKL treatment, even for 
massless quarks.

Before proceeding we should summarise the experimental analyses. 
The ZEUS data are for the energy range 80 GeV$<W_{\gamma p}<100\,\text{GeV}$ 
with the mean value approximately central. Their measurements are 
for 1 GeV$^2 < -t < 10$ GeV$^2$ and they make the cut
$-(M_X^2-m_p^2)/ t < 100 $ 
where $M_X$ is the mass of the proton dissociation products and $m_p$ is the 
mass of the proton. The relation
\begin{eqnarray}
M_X^2-m_p^2=-t(1/x-1)
\end{eqnarray}  
translates this to a cut of $x>0.01$ where $x$ is the longitudinal momentum
fraction of the struck parton. 
ZEUS quotes a photon virtuality $Q^2<0.02\,\text{GeV}^2$ with 
$\langle Q^2\rangle\sim 10^{-5} \,\text{GeV}^2$. This allows us to 
neglect $Q^2$ in the kinematic relations and in our theoretical calculation. 

The H1 data are for $J/\psi$ production, and in the energy range 
50 GeV $<W_{\gamma p}<150\,\text{GeV}$ with the mean value approximately 
central. Their measurements are for 2 GeV$^2 <-t<30\,\text{GeV}^2$ and they
cut in the `elasticity' variable $z_e$, which is related to $M_X$ to a good 
approximation (for the $J/\psi$) by
\begin{eqnarray}
z_e \simeq 1-(M_X^2-t)/W_{\gamma p}^2.
\end{eqnarray}
H1 quote $M_X<30\,\text{GeV}$ which translates to $z_e > 0.95$ and 
$x >  -t/(1-z_e)s$. The photon virtuality is restricted to satisfy 
$Q^2<1\,\text{GeV}^2$ with $\langle Q^2 \rangle \sim 0.06 \,\text{GeV}^2$, 
again small enough to regard the photon as being real. 

Following \cite{FR}, we write the cross-section as a product of the parton
level cross-section and parton distribution functions:
\begin{align}
\frac{d\sigma (\gamma p \rightarrow VX)}{dt\, dx} \;=\;
\biggl(
\frac{4N_c^{4}}{(N_c^{2}-1)^2}
%\left( {C_A \over C_F} \right) ^2
G(x,t)+
\sum_{f}[q_{f}(x,t)+\bar{q}_{f}(x,t)]\biggr)\;
\frac{d\sigma (\gamma q \rightarrow Vq)}{dt},
\label{dsdtgp}
\end{align}
where $N_c=3$, $G(x,t)$ and $q_{f}(x,t)$ are the gluon and quark
distribution functions respectively and $s$ is the $\gamma p$
centre-of-mass energy squared.
The struck parton in the proton, that initiates a jet in the proton
hemisphere, carries a fraction $x$ of the longitudinal momentum
of the incoming proton.

The parton level cross-section, characterised by the invariant
collision energy squared $\hat s = x s$, is expressed in
terms of the helicity amplitudes $M_{\lambda\lambda'} (\hat{s},t)$
($\lambda$ is the photon helicity and $\lambda'$ is the meson helicity):
\begin{equation}
\frac{d \sigma}{dt} = \frac{1}{16 \pi \hat{s}^2}
\left( |M_{++} (\hat s,t)|^2 +
       |M_{+0} (\hat s,t)|^2 +
       |M_{+-} (\hat s,t)|^2 \right)
\label{dsdt}
\end{equation}
The experimenters also measure the spin density matrix elements
$r^{04}_{ij}$ which are extracted from the angular distribution of the
decay products of the vector meson:
\begin{eqnarray}
&&\frac{d^2\sigma}{d\cos \theta_h\,d\phi_h}\propto \biggl[ \frac{1}{2}(1\mp r^{04}_{00})\pm \frac{1}{2}(3r^{04}_{00}-1)\cos^2\theta_h \nonumber \\ &&
\mp \sqrt{2} \text{Re}[r^{04}_{00}]\,\sin 2\theta_h\,\cos \phi_h \mp r^{04}_{1-1}\,\sin^2 \theta_h\,\cos 2\phi_h\biggr],
\label{Heraangdist}
\end{eqnarray}
where the upper (lower) signs are for spin-0 (spin-1/2) particles. 
$\theta_h$ and $\phi_h$ are the spherical polar angles of the 
positive particle of the two body decay of the meson, 
where the vector meson momentum 
defines the $z$-axis and the $x$-axis (which fixes $\phi_h=0$) is defined 
to lie in the direction of the hard momentum transfer, ${\bf q}$. Note
that $|{\bf q}|^2 = -t$ in what follows. 
The $r$-matrix elements can be written \cite{Critt,SSW,SW}

\be
r^{04} _{00} = {\langle |M_{+0}|^2 \rangle \over
        \langle |M_{++}|^2 + |M_{+0}|^2 + |M_{+-}|^2 \rangle },
\label{r00}
\ee

\be
r^{04}_{10} = {1\over 2}
         {\langle M_{++}M^*_{+0} + M_{+-}M^*_{-0} \rangle \over
         \langle |M_{++}|^2 + |M_{+0}|^2 + |M_{+-}|^2 \rangle },
\label{r10}
\ee

\be
r^{04} _{1-1} =\frac{1}{2}{\langle  M_{++}M_{+-}^* +M_{+-}M_{++}^*  \rangle
        \over
         \langle |M_{++}|^2 + |M_{+0}|^2 + |M_{+-}|^2 \rangle },
\label{r1-1}
\ee
where $\langle ... \rangle$ denotes the integration of the
parton level quantities over partonic $x$ with the appropriate
cuts. Throughout this paper we choose a fixed value for the $\gamma p$ 
centre-of-mass energy: $\surd s=W_{\gamma p}=100\,\text{GeV}$.

\section{Theoretical Results}
We begin by summarising the results of \cite{EFMP} for the relevant matrix
elements. We append a superscript to specifiy if the photon-quark
coupling is chiral even or chiral odd.

We use the meson distribution amplitudes of \cite{BBKT,BB} and
use their notation throughout. The explicit forms for the distribution
amplitudes we use can be found in Appendix A. In what follows
we systematically include terms of progressively higher twist in the
distribution amplitudes in order to ascertain their relative importance.
The relevant amplitudes are

\begin{eqnarray}
&& {M}_{+0}^{even}= \frac{isC_V f_V}{4\sqrt{2}|q|}
\int_0 ^1 du \;
 (1-2u)\phi_{\|}(u)
\nonumber \\ 
&& \sum_{n=-\infty}^{n=+\infty}
\int_{-\infty}^{\infty}d\nu
\frac{\nu^{2}+n^{2}}{[\nu^{2}+(n-1/2)^{2}][\nu^{2}+(n+1/2)^{2}]}
\frac{\exp [\chi_{2n}(\nu)z]}{\sin (i\pi\nu)} \,
I_{0-1}(\nu,2n, q, u;1), \nonumber \\ \label{+0ev}
\end{eqnarray}
\begin{eqnarray}
&& {M}_{++}^{even}=\frac{sC_V f_V M_V}{8|q|}\,
\int_0 ^1 du\;
 \left(\frac{g_\perp^{(a)}(u)}{4}-(1-2u)
\int_0^udv(\phi_\|(v)-g_\perp^{(v)}(v)) \right)
\nonumber\\
&& \times \sum_{n=-\infty}^{n=+\infty}
\int_{-\infty}^{\infty}d\nu
\frac{\nu^{2}+n^{2}}{[\nu^{2}+(n-1/2)^{2}][\nu^{2}+(n+1/2)^{2}]}
\frac{\exp [\chi_{2n}(\nu)z]}{\sin (i\pi\nu)} \,
I_{00}(\nu,2n,q, u;1), \nonumber \\ \label{++ev}
\end{eqnarray}
\begin{eqnarray}
&& {M}_{+-}^{even}=
\frac{sC_V f_V M_V}{ 8|q|} \int_0 ^1 du\, 
 \left(\frac{g_\perp^{(a)}(u)}{4}+(1-2u)
\int_0^udv(\phi_\|(v)-g_\perp^{(v)}(v)) \right)
\nonumber \\
&& \times \sum_{n=-\infty}^{n=+\infty}\int_{-\infty}^{\infty}d\nu
\frac{\nu^{2}+n^{2}}{[\nu^{2}+(n-1/2)^{2}][\nu^{2}+(n+1/2)^{2}]}
\frac{\exp [\chi_{2n}(\nu)z]}{\sin (i\pi\nu)} \,
I_{1-1}(\nu,2n, q, u;1). \nonumber \\ \label{+-ev}
\end{eqnarray}
\begin{eqnarray}
&& {M}_{+0}^{odd}=
\frac{isC_V f_V ^T M_V}{4\sqrt{2}|q|}
\int_0 ^1 du \;
\int_0^udv(h_\|^{(t)}(v)-\phi_\perp(v))
\nonumber \\
&&
\times \sum_{n=-\infty}^{n=+\infty}
\int_{-\infty}^{\infty}d\nu
\frac{\nu^{2}+n^{2}}{[\nu^{2}+(n-1/2)^{2}][\nu^{2}+(n+1/2)^{2}]}
\frac{\exp [\chi_{2n}(\nu)z]}{\sin (i\pi\nu)} \,
I_{{1 \over 2}\, -{1\over 2}}(\nu,2n, q, u;0),
\nonumber\\
\label{+0odd}
\end{eqnarray}
\begin{eqnarray}
&& {M}_{++}^{odd}=\frac{sC_V f_V^T}{4|q|}\,
\int_0 ^1 du\;
\phi_{\perp}(u)\,
%u\bar{u}\biggl( \int_{0}^{u}\frac{dv}{\bar{v}}\phi_{\parallel}(v)+
%\int_{u}^{1}\frac{dv}{v}\phi_{\parallel}(v)\biggr)
\nonumber\\
&& \times \sum_{n=-\infty}^{n=+\infty}
\int_{-\infty}^{\infty}d\nu
\frac{\nu^{2}+n^{2}}{[\nu^{2}+(n-1/2)^{2}][\nu^{2}+(n+1/2)^{2}]}
\frac{\exp [\chi_{2n}(\nu)z]}{\sin (i\pi\nu)} \,
I_{-{1\over 2}\, -{1\over 2}}(\nu,2n,q, u;0),
\nonumber
\\
\label{++odd}
\end{eqnarray}
\begin{eqnarray}
&& {M}_{+-}^{odd}=
\frac{sC_V f_V ^T M_V^2}{ 8|q|} \int_0 ^1 du\,
%\biggl(\bar{u}^{2} \int_{0}^{u}\frac{dv}{\bar{v}}\phi_{\parallel}(v)+u^{2}
%\int_{u}^{1}\frac{dv}{v}\phi_{\parallel}(v)\biggr)\,
 \int_0^u dv\int_0^vd\eta\left(h_{\|}^{(t)}(\eta)-\frac{1}{2}\phi_{\perp}(\eta)-\frac{1}{2}h_3(\eta)\right)
\nonumber \\
&& \times \sum_{n=-\infty}^{n=+\infty}\int_{-\infty}^{\infty}d\nu
\frac{\nu^{2}+n^{2}}{[\nu^{2}+(n-1/2)^{2}][\nu^{2}+(n+1/2)^{2}]}
\frac{\exp [\chi_{2n}(\nu)z]}{\sin (i\pi\nu)} \,
I_{{3\over 2}\, -{1\over 2}}(\nu,2n, q, u;0).
\nonumber
\\
\label{+-odd}
\end{eqnarray}
where
\ba
I_{\alpha\beta}(\nu, n, q, u;a) &=&
\mq \int  d^{2}\rho\;\rho^{\alpha+1} \rho^{*\,\beta+1}\, K_a(\mq |\rho|)
e^{\frac{i\xi}{4}[q^{*}\rho + q \rho^{*} ]}  \nonumber
\\
&\times& [J_{\mu}(q^{*}\rho/4) J_{\tilde\mu}(q\rho^{*}/4) -
(-1)^n \; J_{-\mu}(q^{*}\rho/4) J_{-\tilde\mu}(q\rho^{*}/4)],
\label{ialbe}
\ea
$\xi=2u-1$, $\mu=n/2-i \nu$, $\tilde{\mu}=-n/2-i \nu$, 
$K_a(x)$ is the modified Bessel
function and the parameter $a$ equals $1$ for the chiral-even
and $0$ for the chiral-odd contributions. In (\ref{ialbe}) we use
complex variable notation for the transverse vectors $\rho$ and $q$, e.g. 
$\rho = \rho_x + i \rho_y$.

After some effort \cite{EFMP} one finds
\ba
I_{\alpha\beta}(\nu,n,q,u;a) &=&
\frac{\mq}{2}\int^{C^{\prime}+i\infty}_{C^{\prime}-i\infty}
\frac{d\zeta}{2\pi i}\Gamma(a/2-\zeta)\Gamma(-a/2-\zeta)\,
\tau_q ^{\zeta} \; (i\, \text{sign}\,(1-2u))^{\alpha-\beta+n}  \nonumber \\
&\times&  \left(\frac{4}{|q|}\right)^{4}
\left[\sin\pi(\alpha + \mu + \zeta)\; \right.
B(\alpha,\mu, q^* ,u,\zeta)\,
B(\beta,\widetilde\mu,q,u^* ,\zeta) \nonumber \\
&-& (-1)^n
\sin\pi(\alpha - \mu + \zeta)\;
B(\alpha,-\mu, q^* ,u,\zeta)\,
B(\beta,-\widetilde\mu,q,u^* ,\zeta)
\left. \right]
\label{finalintegral}
\ea
where we have introduced the dimensionless parameter $\tau_q = 4\mq^2/|q|^2$
and the conformal blocks
\ba
B(\alpha,\mu, q^* ,u,\zeta) =  
(-4u \bar u)^{-(\mu+2+\alpha+\zeta)/2}
\left(\frac{4}{ q^* }\right)^\alpha
2^{-\mu}\,
\frac{\Gamma(\mu+2+\alpha+\zeta)}{\Gamma(\mu+1)} \nonumber \\
\f\left(\frac{\mu+2+\alpha+\zeta}{2} \, , \,
\frac{\mu-1-\alpha-\zeta}{2}\, ; \,
\mu+1\, ; \,\frac{1}{4u \bar u}\right).
\label{blocks2}
\ea
Note, that the sums are performed over even
conformal spins $2n$.

We have also introduced the energy variable
\begin{eqnarray}
z=\frac{3\alpha_s}{2\pi}\ln \left( \frac{xs}{\Lambda^2}\right)
\label{zrap}
\end{eqnarray}
where $\Lambda$ is an undetermined energy scale in the leading logarithmic
approximation. The eigenvalues of the BFKL kernel are proportional to 
\begin{equation}
\chi_{n}(\nu)\;=\; 4\repart \biggl(\psi(1)-\psi(1/2+|n|/2+i\nu)\biggr)
\end{equation}
\begin{table}
\begin{center}
\begin{tabular}{|c||c|c|c|} \hline
 & $\rho$ & $\phi$ & $J/\psi$ \\ \hline
$Q_V$  & $1/\sqrt{2}$  & $-1/3$ & $2/3$ \\ \hline
\end{tabular}
\end{center}
\caption{The meson electric charge couplings.}
\label{Qv}
\end{table}
and we collect together several factors into $C_V$:
\begin{equation}
C_V = i \alpha_s^2 \frac{N_c^2-1}{N_c^2} e Q_V
\end{equation}
where $Q_V$ is determined by the electric charge of the quarks which make
up the meson. The explicit values are listed in Table \ref{Qv}.

All the helicity amplitudes depend upon the vector meson coupling constants; 
either $f_V$ (vector) or $f_V^T$ (tensor). Table \ref{f} presents the 
vector and tensor couplings for the relevant mesons. The vector 
couplings are directly measurable from the electronic decay width of the 
vector mesons whilst the tensor couplings have been 
calculated at $\mu=1\,\text{GeV}$ using QCD sum rules 
\cite{BBKT}\footnote{The tensor coupling for the $J/\psi$ value is 
unavailable in the literature and so we, as default, put it equal to the 
vector coupling.}.

\begin{table}
\begin{center}
\begin{tabular}{|c||c|c|c|} \hline
 & $\rho$ & $\phi$ & $J/\psi$ \\ \hline
$f_V$ [GeV] & $0.216$ & $0.231$ & 0.405 \\ \hline
$f_V^T(1\,\text{GeV})$ [GeV] & $0.160$ & $0.215$ & (0.405) \\ \hline
\end{tabular}
\end{center}
\caption{The meson decay constants \cite{BBKT}.}
\label{f}
\end{table}

In Table \ref{HA} we classify the above set of helicity amplitudes in terms
of twist. We work to next-to-leading twist in each amplitude, i.e. we
do not consider twist-4 terms in the single and no flip amplitudes.

\begin{table}
\begin{center}
\begin{tabular}{|c|c|c|c|} \hline
twist & 2 & 3 & 4  \\ \hline \hline
single-flip & ${M}_{+0}^{even}$ & ${M}_{+0}^{odd}$ & $-$ \\ \hline
no-flip & ${M}_{++}^{odd}$ & ${M}_{++}^{even}$ & $-$ \\ \hline
double-flip & $0$ & ${M}_{+-}^{even}$ & ${M}_{+-}^{odd}$ \\ \hline
\end{tabular}
\end{center}
\caption{Classification of the relevant helicity amplitudes in twist.}
\label{HA}
\end{table}

\section{Parameters and prescriptions}

Strictly, in perturbation theory, we should use the 
current quark mass.  However, using the current mass renders the chiral odd 
contribution negligible for light mesons and proves to be incapable of 
describing the $r$-matrix elements.  The authors of \cite{IKSS} noted 
the importance of introducing a large chiral odd contribution. The method 
they employ introduces a large non-perturbative coupling of the photon to
chiral odd quark-antiquark configurations. We shall take a 
different approach and replace the current 
quark mass with the constituent quark mass, i.e. we take $m=m_V/2$. 
We will investigate the sensitivity to this parameter.

We also need to explain
how we implement the strong coupling, $\alpha_s$ and the scale $\Lambda^2$
which appears in (\ref{zrap}). We might reasonably expect that  
$\Lambda^2=\beta m_V^2 - \gamma t$
where $\beta$ and $\gamma$ are unknown. To simplify matters, 
we choose to fix, $\beta=1$ since a change in its value 
can approximately be traded off with a change in $\alpha_s$. 

Notice that $\alpha_s$ appears in two places. It appears 
as a prefactor of $\alpha_s^2$ in every helicity amplitude. 
It also appears in the definition of $z$ (see (\ref{zrap})). 
The prefactor arises from the coupling of the two gluons 
to each impact factor. The coupling in $z$, however, is generated by the 
gluon couplings inside the gluon ladder. We shall denote these
couplings $\alpha_s^{IF}$ and $\alpha_s^{BFKL}$ respectively. 
Strictly speaking a fixed value of $\alpha_s$ is appropriate to the level 
of our calculation. However, next-to-leading logarithms cause $\alpha_s$  
to run. Arguments exist that the effect of these corrections is to 
effectively fix $\alpha_s^{BFKL}$ \cite{BFKLP}. Phenomenologically, fixing 
$\alpha_s^{BFKL}$ has proved successful in the past \cite{FP,CFL} and we will 
fix it here. We remain free to run or fix the impact factor coupling, 
$\alpha_s^{IF}$, and we shall subsequently make use of this possibility. 
When we refer to a running coupling in future, we shall be referring to
its value at $1\,\text{GeV}$. 
We note that, as ratios, the $r$-matrix elements are independent of 
$\alpha_s^{IF}$. 

In summary, the parameters we shall treat as free for fitting purposes are
\begin{eqnarray}
\alpha_s^{IF},\hspace{1cm}\alpha_s^{BFKL}\hspace{1cm}\text{and}\hspace{1cm} 
\Lambda^2=m_V^2-\gamma t. 
\end{eqnarray}
We shall also show the sensitivity of our variables to the quark mass
$m$ though we shall not treat this as a fit parameter.

For other parameters, we shall use either those measured by experiment or 
estimated in the literature. We shall not vary these. The sensitivity 
of our predictions to various approximations for the meson distribution
amplitudes will be tested by testing the importance of $1/|{\bf q}|^2$ 
suppressed and above terms and by employing three different distribution 
amplitude prescriptions. We refer to the three prescriptions as the 
$\delta$-function prescription, the asymptotic distribution amplitude 
prescription (obtained in the limit $\alpha_s\rightarrow 0$) and the 
full distribution amplitude prescription (of \cite{BBKT,BB}). 
All necessary information on the meson distribution amplitudes can be found in 
Appendix \ref{DistA}.

\section{Phenomenology for the $\rho$ meson}

In this section we focus wholly on the phenomenology of the $\rho$ meson.
We begin with the most crude approximations for the meson wavefunction
and systematically relax these to the point where we eventually treat the
meson using the full light-cone wavefunctions of \cite{BBKT,BB}.

\subsection{The collinear and $\delta$-function approximations}

The collinear approximation corresponds to putting the transverse momentum 
of the quark and antiquark exiting the hard scatter (measured relative
to the meson direction) to zero. This approximation
generates $\delta$-functions which force the transverse size of the 
quark-antiquark pair to zero. This fact forces all the higher twist 
helicity amplitudes to zero so that only ${M}_{+0}^{even}$ 
and ${M}_{++}^{odd}$ (twist-2) then survive.

The $\delta$-function approximation for the meson distribution amplitudes 
corresponds to equating
\begin{eqnarray}
\phi_{\|}(u)=\phi_{\perp}(u)=\delta(u-1/2).
\end{eqnarray}
The ${M}_{+0}^{even}$ helicity amplitude is zero 
in this approximation and so the only non-zero amplitude is 
${M}_{++}^{odd}$:
\begin{eqnarray}
{M}_{++}^{odd}=sC_V\frac{8\pi}{t^2}\,f_Vm_V\,\hspace{6cm}
\nonumber \\ \times
\sum_{m=-\infty}^{m=\infty}
\biggl( -\frac{1}{4}\biggr)^{|m|}\int \,d\nu\,
\frac{\nu^{2}+m^{2}}{(\nu^{2}+(m-1/2)^{2})(\nu^{2}+(m+1/2)^{2})}
\; e^{\chi_m(\nu)z} \
\nonumber \\
\times
\int_{-i\infty}^{i\infty}\frac{ds}{2\pi i}\, \tau^{1/2+s+|m|}\,
\frac{\Gamma(1-s-i\nu)\,\Gamma(1-s+i\nu)}
     {\Gamma(1-s/2-i\nu /2)\,\Gamma(1-s/2+i\nu /2)}
\nonumber \\
\times
\frac{\Gamma^{2}(1/2+s+|m|)}
{\Gamma(1/2+s/2-i\nu/2+|m|)\,\Gamma(1/2+s/2+i\nu /2+|m|)}.
\label{hvm}
\end{eqnarray}

\begin{figure}
{\par\centering {\includegraphics[height=10.5cm,angle=-90]{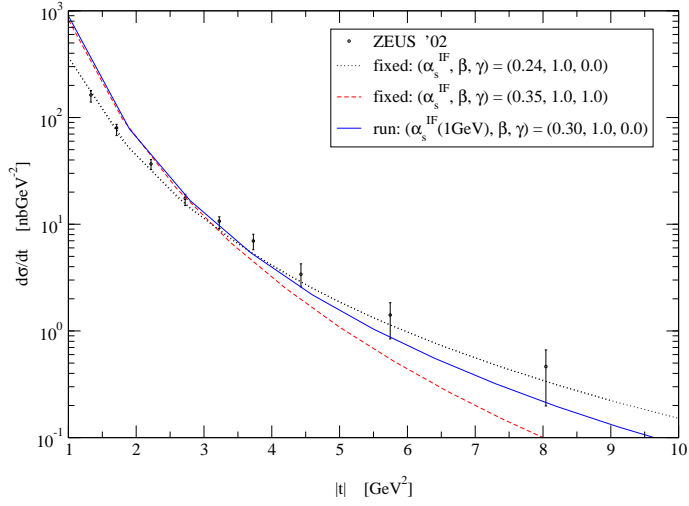}} \par}
\caption{Collinear and $\delta$-function approximation: $d\sigma/dt$ for parameter values, 
 $(\alpha_s^{BFKL},\,m) = (0.2,\,m_\rho/2)$.}
\label{stcoldel}
\end{figure}

Figure \ref{stcoldel} shows the predictions for $d\sigma/dt$ for three 
different sets of parameter values. Running the scale $\Lambda^2$ that 
appears under $s$ in the BFKL logarithm (\ref{zrap}) with $t$ or 
running $\alpha_s^{IF}$ 
with $t$ prove to predict $t$ dependencies that are too steep and 
incompatible with the data presented by ZEUS \cite{ZEUS1}. Fixing the strong 
coupling in the impact factor and in the BFKL ladder in addition to the 
BFKL logarithm yields a good fit. These results agree with those of \cite{FP} 
for the leading conformal spin component ($n=0$)\footnote{Note that in 
\cite{FP}, the authors implicitly equated the tensor vector meson 
coupling, $f_V^T$, to the vector coupling, $f_V$, rather than using the 
value estimated from QCD sum rules.} and \cite{HVM} for $n>0$.

\begin{figure}
{\par\centering {\includegraphics[height=10.5cm,angle=-90]{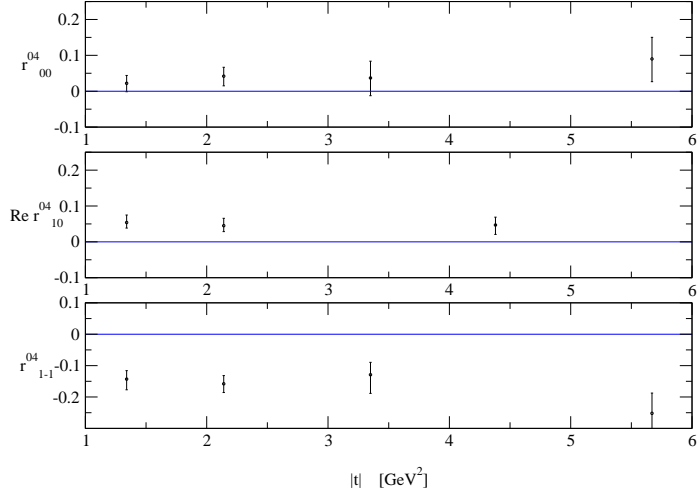}} \par}
\caption{Collinear and $\delta$-function approximation: $r$-matrix elements. The predictions are parameter independent.}
\label{sdmcoldel}
\end{figure}

Since only the ${M}_{++}^{odd}$ amplitude is non-zero with the 
collinear and $\delta$-function approximations imposed, referring to their 
definitions ((\ref{r00}), (\ref{r10}), (\ref{r1-1})), we see that this 
enforces all the $r$-matrix elements to zero. Figure \ref{sdmcoldel} 
shows this prediction in comparison to the ZEUS data. We see that zero 
$r$-matrix elements are compatible with data for $r^{04}_{00}$, in fair 
agreement for $r^{04}_{10}$ and in poor agreement for $r^{04}_{1-1}$.

\subsection{Collinear approximation and asymptotic distribution amplitudes}

The use of a $\delta$-function distribution amplitude is a crude 
approximation. For the charm quark and $J/\psi$ meson, 
$m_c\sim m_{J/\psi}/2$. For this meson, the configurations that provide 
the dominant contributions should be close to $u\sim 1/2$, i.e. we expect 
the  $\delta$-function approximation to be fair. This is not so for the 
light $\rho$. The use of the $\delta$-function was motivated purely by past 
phenomenological success \cite{FP}. To understand 
this success, and to have any hope of predicting the helicity structure of 
the cross-section, we must relax this crude assumption. 
For now, we maintain the collinear assumption and use the 
asymptotic distribution 
amplitudes. The relevant amplitudes are
\begin{eqnarray}
\phi_{\|}(u)=\phi_{\perp}(u)=6u(1-u).
\end{eqnarray} 
Now momentum configurations of $u\not= 1/2$ contribute and the 
${M}_{+0}^{even}$ helicity amplitude is no longer zero.

\begin{figure}
{\par\centering {\includegraphics[height=10.5cm,angle=-90]{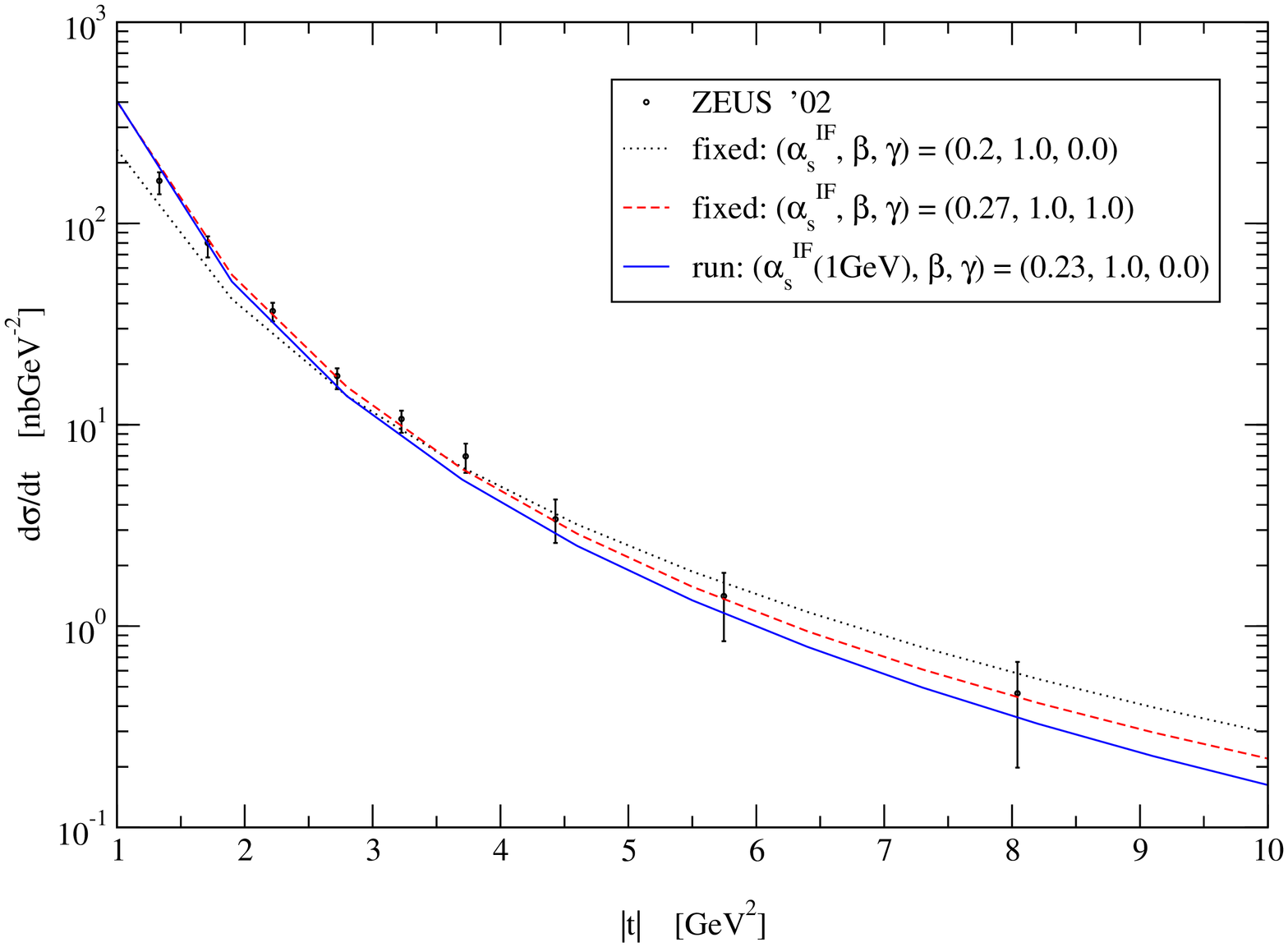}} \par}
\caption{Collinear approximation: $d\sigma/dt$ for parameter values, 
 $(\alpha_s^{BFKL},\,m) = (0.2,\,m_\rho/2)$.}
\label{stcol}
\end{figure}

\begin{figure}

{\par\centering {\includegraphics[height=10.5cm,angle=-90]{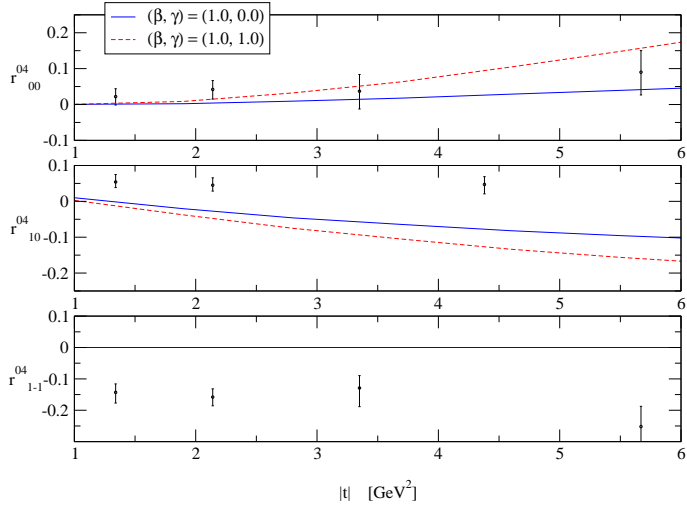}} \par}
\caption{Collinear approximation: $r$-matrix elements for parameter values, 
 $(\alpha_s^{BFKL},\,m)$ $ =$ $ (0.2,\,m_\rho/2)$.}
\label{sdmcol}

\end{figure}

Figure \ref{stcol} shows predictions for the cross-section. 
The ${M}_{+0}^{even}$ amplitude has a less steep $t$ dependence 
than ${M}_{++}^{odd}$. The higher suppression of the 
chiral odd term is what we expect since in the definition of 
the $I_{\alpha\beta}$ integral in ${\bf r}$ space (see (\ref{ialbe})), 
the chiral even amplitudes have a $K_1$ function and the chiral odd 
ones a $K_0$. The $r^{04}_{00}$ element is 
the ratio of the longitudinal component of the cross-section to the total. 
Figure \ref{sdmcol} demonstrates that this longitudinal component is 
numerically subdominant, but becomes increasingly important with rising 
$|t|$. The change in the $t$ dependence of the combined cross-section, 
due to this component, results in the previous fit of $\gamma=0$ and 
fixed $\alpha_s^{IF}$ now being marginal. We now fit the data better by 
either running $\Lambda^2$ with $t$ or $\alpha_s^{IF}$ with $t$.

Note that the amplitude ${M}_{+-}$ is still zero in our collinear 
scheme. This means that the $r^{04}_{1-1}$ $r$-matrix element is still 
predicted to be zero. As observed for the case of the $\delta$-function 
distribution amplitude, this is precisely the $r$-matrix most incompatible 
with such a prediction. Figure \ref{sdmcol} shows the effect on the 
$r$-matrix elements of running the $\Lambda^2$ scale with $t$ versus 
keeping it fixed. We observe that both provide a good fit for 
$r^{04}_{00}$ and poor predictions for $r^{04}_{10}$ and 
$r^{04}_{1-1}$\footnote{Note that a flip in sign of 
either ${M}_{+0}$ or both ${M}_{++}$ and ${M}_{+-}$ 
would flip the sign of our prediction for $r^{04}_{00}$. We note that
our two-gluon leading twist amplitudes agree with those of \cite{GPS1} 
and we checked that our BFKL result reduces to the two-gluon prediction 
in the limit $z\rightarrow 0$. In addition we have checked that all our 
conventions agree with those of ZEUS, i.e. the direction of 
${\bf q}$, and that we have the correct 
phases for the polarisation vectors.}. 
The poor quality of fits to the $r$-matrix elements 
motivates the progression to higher twist. Higher twist is the only 
solution to producing a non-zero ${M}_{+-}$ contribution.   


\subsection{Higher twist effects}

Higher twist effects are introduced by relaxing the assumption that a 
collinear quark and antiquark exit the hard subprocess (i.e. 
we no longer have dipoles of zero transverse size) or by inclusion of 
higher Fock states (e.g.\ $q\bar q g$) in the wave function.

Potentially, we must consider the full six helicity amplitudes. The 
complete set of relevant asymptotic distribution amplitudes are
\begin{eqnarray}
&& \phi_{\|}(u)=\phi_\perp (u)=6u\bar{u} \nonumber \\ &&
g_{\perp}^{(a)}(u)=6u\bar{u}\hspace{1cm} g_{\perp}^{(v)}(u)=\frac{3}{4}(1+\xi^2)
\nonumber \\ &&
h_3(u)=\frac{3}{2}(1-\xi^2),
\end{eqnarray}
where $\xi=2u-1$.
Substituting these into (\ref{+0ev})--(\ref{+-odd}) 
we obtain the amplitudes which were presented
explicitly in \cite{EFMP}. We do not list them again here.

We want to explore the effect of higher twist terms on our observables by 
systematically adding suppressed terms. 
Any observable is proportional to products of the helicity amplitudes 
i.e. ${M}_{ij}{M}^{*}_{kl}$. Consider the (parton level) 
cross-section for the production of a longitudinal meson:
\begin{eqnarray}
&& \hspace{1.5cm}\frac{d\sigma_{+0}}{dt}\propto 
{M}_{+0}{M}^{*}_{+0}= \biggl({M}_{+0}^{even}+{M}_{+0}^{odd}
\biggr)\biggl({M}_{+0}^{even\,*}+{M}_{+0}^{odd\,*}\biggr)  \nonumber \\ &&
={M}_{+0}^{even}{M}_{+0}^{even\,*}+{M}_{+0}^{even}{M}_{+0}^{odd\,*}+
{M}_{+0}^{odd}{M}_{+0}^{even\,*}+{M}_{+0}^{odd}{M}_{+0}^{odd\,*} 
\nonumber \\ &&
\hspace{0.3cm}{}^\text{(twist-2 x twist-2)}\hspace{0.3cm} 
{}^\text{(twist-2 x twist-3)} \hspace{0.3cm}{}^\text{(twist-2 x twist-3)}
\hspace{0.3cm}{}^\text{(twist-3 x twist-3)}\nonumber
\end{eqnarray}
In the expression above we see that the first term is leading  
(twist-2 x twist-2), the second and third $1/|{\bf q}|$ 
suppressed (twist-2 x twist-3), and the fourth $1/|{\bf q}|^2$ 
suppressed (twist-3 x twist-3). Note that a (twist-2 x twist-4) term 
would also be $1/|{\bf q}|^2$ suppressed. Next-to-leading power behaviour 
in observables is therefore the highest level in which we are able to 
compute. Note that the helicity double-flip amplitude is subleading 
at this level of approximation as far as the total cross-section is 
concerned but that it provides the leading behaviour for $r_{1-1}^{04}$.

\begin{figure}
{\par\centering {\includegraphics[height=10.5cm,angle=-90]{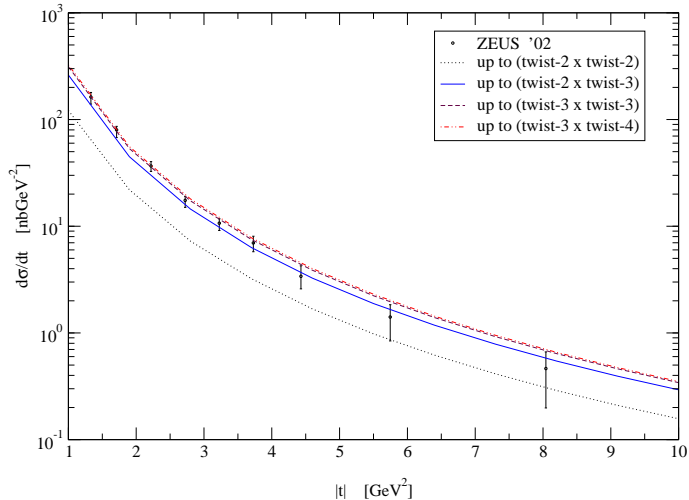}} \par}
\caption{Higher twist effects: $d\sigma/dt$ for parameter values, 
 $(\alpha_s^{IF},\alpha_s^{BFKL},\,\,\Lambda^2,\,m) = (0.17,\,0.2,\,m_V^2,\,m_\rho/2)$.}
\label{stht}
\end{figure}

Figure \ref{stht} shows $d\sigma/dt$ for the varying levels of approximation. 
The effect on the combined cross-section is predominantly that of 
normalisation. We see that adding in the (twist-2 x twist-3) to the 
(twist-2 x twist-2) terms has a significant effect, but that adding those 
next-to-next-to-leading terms that we have knowledge of has much less effect. 
This reassuring result suggests that the total cross-section 
is insensitive to the twist-four terms that we have neglected.

\begin{figure}
{\par\centering {\includegraphics[height=10.5cm,angle=-90]{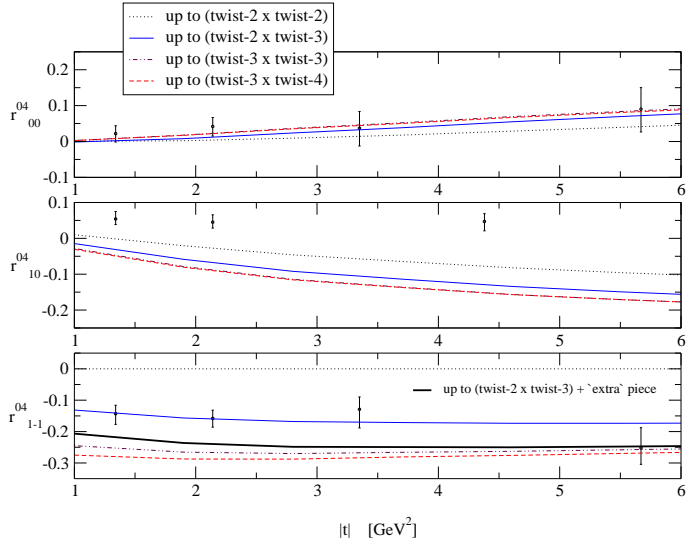}} \par}
\caption{Higher twist effects: $r$-matrix elements for parameter values, 
 $(\alpha_s^{BFKL},\,\Lambda^2,\,m)$ $ =$ $ (0.2,\,m_V^2,\,m_\rho/2)$.}
\label{sdmht}
\end{figure}

\begin{figure}
{\par\centering {\includegraphics[height=10.5cm,angle=-90]{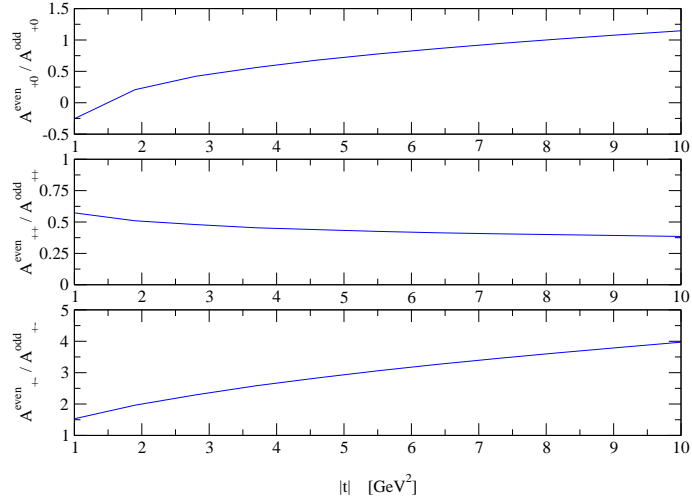}} \par}
\caption{Higher twist effects: even to odd amplitude ratios for parameter values, 
 $(\alpha_s^{BFKL},\,\Lambda^2,\,m) = (0.2,\,m_V^2,\,m_\rho/2)$.}
\label{arht}
\end{figure}

Figure \ref{sdmht} shows that the observables $r^{04}_{00}$ and 
$r^{04}_{10}$ are also relatively insensitive to next-to-next-to leading 
terms. The $r^{04}_{1-1}$ proves more sensitive. Adding the (twist-3 x 
twist-3) terms has an effect of order $100 \%$. It is not surprising that 
this observable is the most sensitive to higher twist, since it is zero at 
leading twist. It therefore appears that we cannot trust our predictions 
for this $r$-matrix element and that it is not a good observable 
with which to test the validity of BFKL dynamics. However, Figure 
\ref{arht}, makes it clearer what is happening. It shows the ratios of 
even to odd helicity amplitudes at $x=0.1$\footnote{At HERA, 
the mean $x$ values that contribute for each amplitude are $\langle x\rangle 
\simeq 0.1$}. Our twist counting scheme ignores the effect 
of $|{\bf q}|$-dependent effects in the hard scatter. The plots in 
Figure \ref{arht} are a test of the validity of this assumption. If 
effects of the hard scatter were the same, we would expect twist 
counting to give us ${M}_{+0}^{even}/{M}_{+0}^{odd}$ 
and  ${M}_{+-}^{even}/{M}_{+-}^{odd}$ approximately 
$\propto \sqrt{-t}$ and  ${M}_{++}^{even}/{M}_{++}^{odd}$ 
$\propto 1/\sqrt{-t}$. Our assumption seems reasonable for the $(+0)$ and 
$(+-)$ amplitudes, however, for $(++)$ the odd and even terms have a 
similar $t$-dependence. ${M}_{++}^{even}$ is twist-3 but the 
${\bf q}$ dependence from the hard scatter boosts its importance to the same 
significance as the twist-2 pieces. The thick line in the plot of 
$r^{04}_{1-1}$ in Figure \ref{sdmht}, is the prediction where we, in 
addition to the  ${M}_{++}^{odd}{M}_{+-}^{even}$ 
term, kept the ${M}_{++}^{even}{M}_{+-}^{even}$ term. 
For $t$ significantly above $1\,\text{GeV}^2$, we can see that the effect 
of terms genuinely $1/|{\bf q}|^2$ suppressed and above, is minimal. Thus, 
though  $r_{1-1}^{04}$ is the most sensitive observable to higher twist 
effects, we can claim to have control over it. 
We shall therefore, from here on, use the 
formulae (\ref{+0ev})--(\ref{+-odd}) with the distribution amplitudes 
computed up to
twist-4 for $M_{+-}$ and up to twist-3 for all others.
Our control over higher twist suggests that higher twist effects 
are unlikely to provide the solution to the fact we cannot fit the complete 
set of $r$-matrix elements.


\subsection{The sensitivity of predictions to parameters}

We demonstrated in the previous section that, away from low $t$, our 
predictions are only weakly sensitive to higher twist corrections. The 
fit for the $r^{04}_{10}$ $r$-matrix element remained particularly poor, 
however. 
In this section we explore the sensitivity of our results to
varying the various parameters at our disposal. 
We shall use a fixed $\alpha_s^{IF}=0.17$, and it should then be
understood that only the shape of the $t$-distribution matters 
(its normalisation being adjustable by varying $\alpha_s^{IF}$). 
Of course $\alpha_s^{IF}$ cancels in the $r_{ij}^{04}$.

\subsubsection{Sensitivity to $\alpha_s^{BFKL}$}

\begin{figure}

{\par\centering {\includegraphics[height=10.5cm,angle=-90]{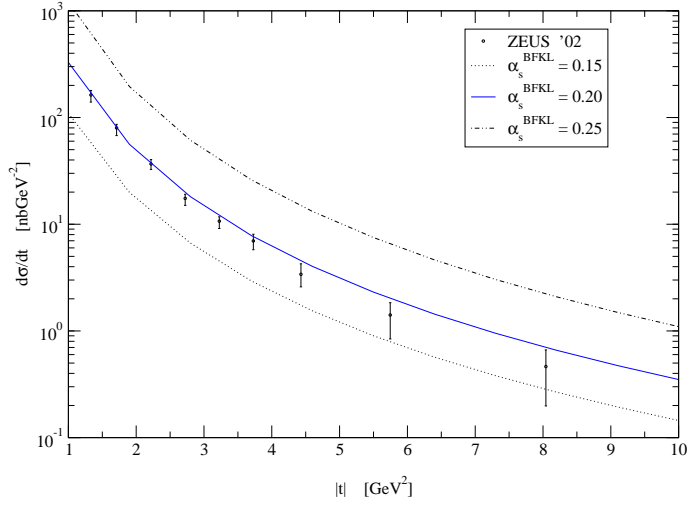}} \par}
\caption{The sensitivity to varying $\alpha_s^{BFKL}$: $d\sigma/dt$ for parameter values, 
 $(\alpha_s^{IF},\,\Lambda^2,\,m)$ $ =$ $(0.17,\,m_\rho^2-t,\,m_\rho/2)$.}
\label{stas}

\end{figure}

Figure \ref{stas} shows a central curve corresponding to: 
$(\alpha_s^{IF},\,\alpha_s^{BFKL},\,\Lambda^2,\,m)$ $=$ \newline $(0.17,\,0.2,\,m_\rho^2,\,m_\rho/2)$. The two other curves show the effect of 
varying the value of $\alpha_s^{BFKL}$ by $25\%$. The effect is 
primarily that of normalisation.

\begin{figure}

{\par\centering {\includegraphics[height=10.5cm,angle=-90]{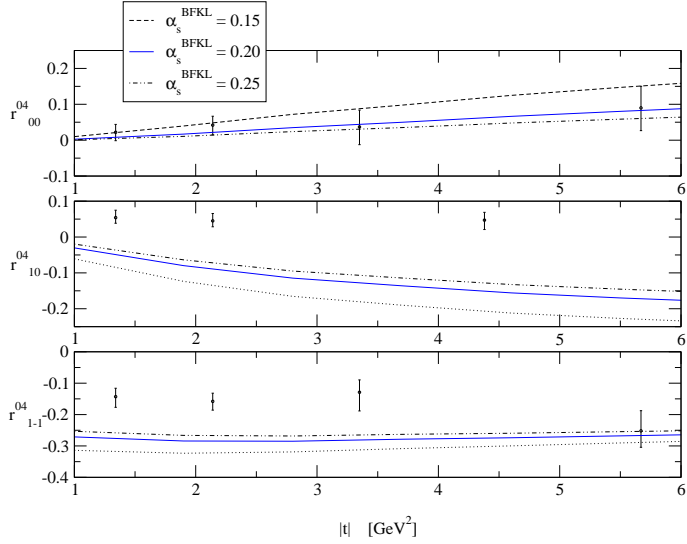}} \par}
\caption{The sensitivity to varying $\alpha_s^{BFKL}$: $r$-matrix elements for parameter values, 
  $(\Lambda^2,\,m)$ $ =$ $(m_\rho^2-t,\,m_\rho/2)$.}
\label{sdmas}

\end{figure}

The $r$-matrix element predictions are shown in Figure \ref{sdmas}. 
$r_{00}^{04}$ is the most sensitive to variations in 
$\alpha_s^{BFKL}$, followed in order by $r_{10}^{04}$ and $r_{1-1}^{04}$. 
We find $r_{1-1}^{04}$ to be relatively insensitive to this parameter 
whereas we remind the reader that it proved most sensitive to higher twist. 
We note that while the general quality of the fits for all the $r$-matrix 
elements is unaffected by varying $\alpha_s^{BFKL}$ by $25\%$, a large 
value of $\alpha_s$ is associated with a larger suppression of the 
longitudinal cross-section. We also note that the fit to $r^{04}_{10}$ 
remains poor.  

\subsubsection{Sensitivity to $\gamma$}

\begin{figure}

{\par\centering {\includegraphics[height=10.5cm,angle=-90]{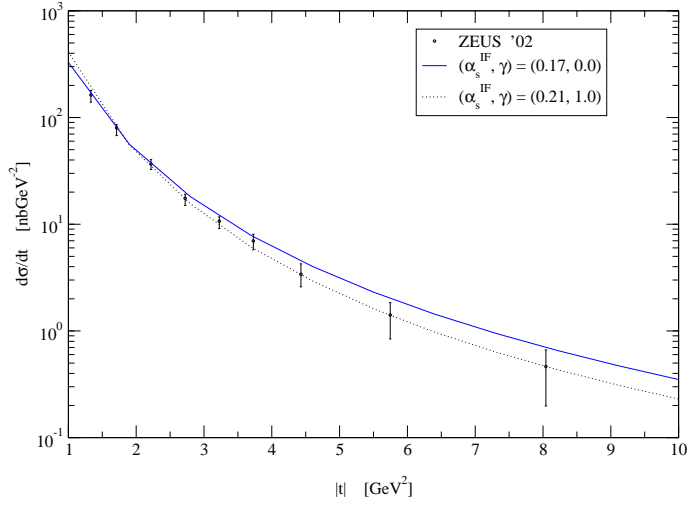}} \par}
\caption{The sensitivity to varying $\gamma$: $d\sigma/dt$ for parameter values, 
 $(\alpha_s^{IF},\,\alpha_s^{BFKL},\,m) = (0.17,\,0.2,\,m_\rho/2)$.}
\label{stgam}
\end{figure}

The default set, 
$(\alpha_s^{IF},\,\alpha_s^{BFKL},\,\Lambda^2,\,m)=(0.17,\,0.2,\,
m_\rho^2,\,m_\rho/2)$, that provides the solid curve in Figure \ref{stgam}, 
is only marginally compatible with the cross-section data. This is 
because, as previously noted, the longitudinal contribution becomes 
increasingly important at higher $t$ and this flattens the $t$ dependence. 
The dotted curve on the same graph demonstrates that we can fit the data 
much better by running the BFKL scale, $\Lambda^2$.

\begin{figure}
{\par\centering {\includegraphics[height=10.5cm,angle=-90]{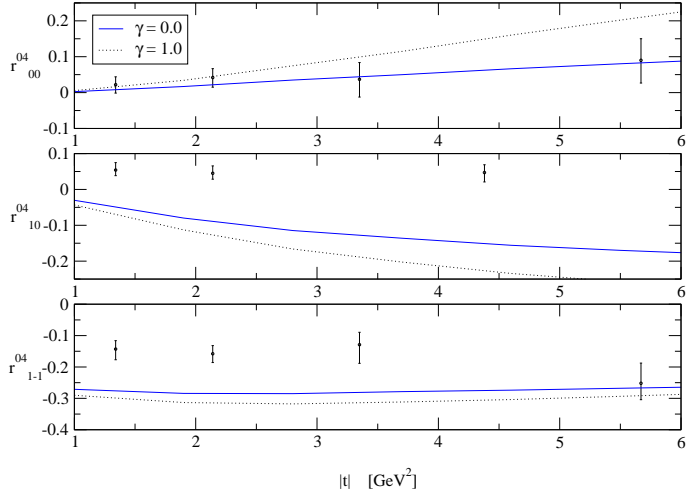}} \par}
\caption{The sensitivity to varying $\gamma$: $r$-matrix elements for parameter values, 
 $(\alpha_s^{BFKL},\,m) = (0.2,\,m_\rho/2)$.}
\label{sdmgam}

\end{figure}

The predictions in Figure \ref{sdmgam} for the $r$-matrix elements again 
show the sign problem in the prediction for $r_{10}^{04}$. Running 
$\Lambda^2$ has not solved this. Note also that the predictions for 
$r_{00}^{04}$ and $r_{10}^{04}$ are sensitive to the values of the BFKL 
parameters and in particular, the quality of the fits becomes worse when 
$\Lambda^2$ is run with $t$. We see that when $\Lambda^2$ is run, the 
longitudinal fraction grows quicker than is compatible with data.

\subsubsection{Sensitivity to $z$ rapidity}

\begin{figure}

{\par\centering {\includegraphics[height=10.5cm,angle=-90]{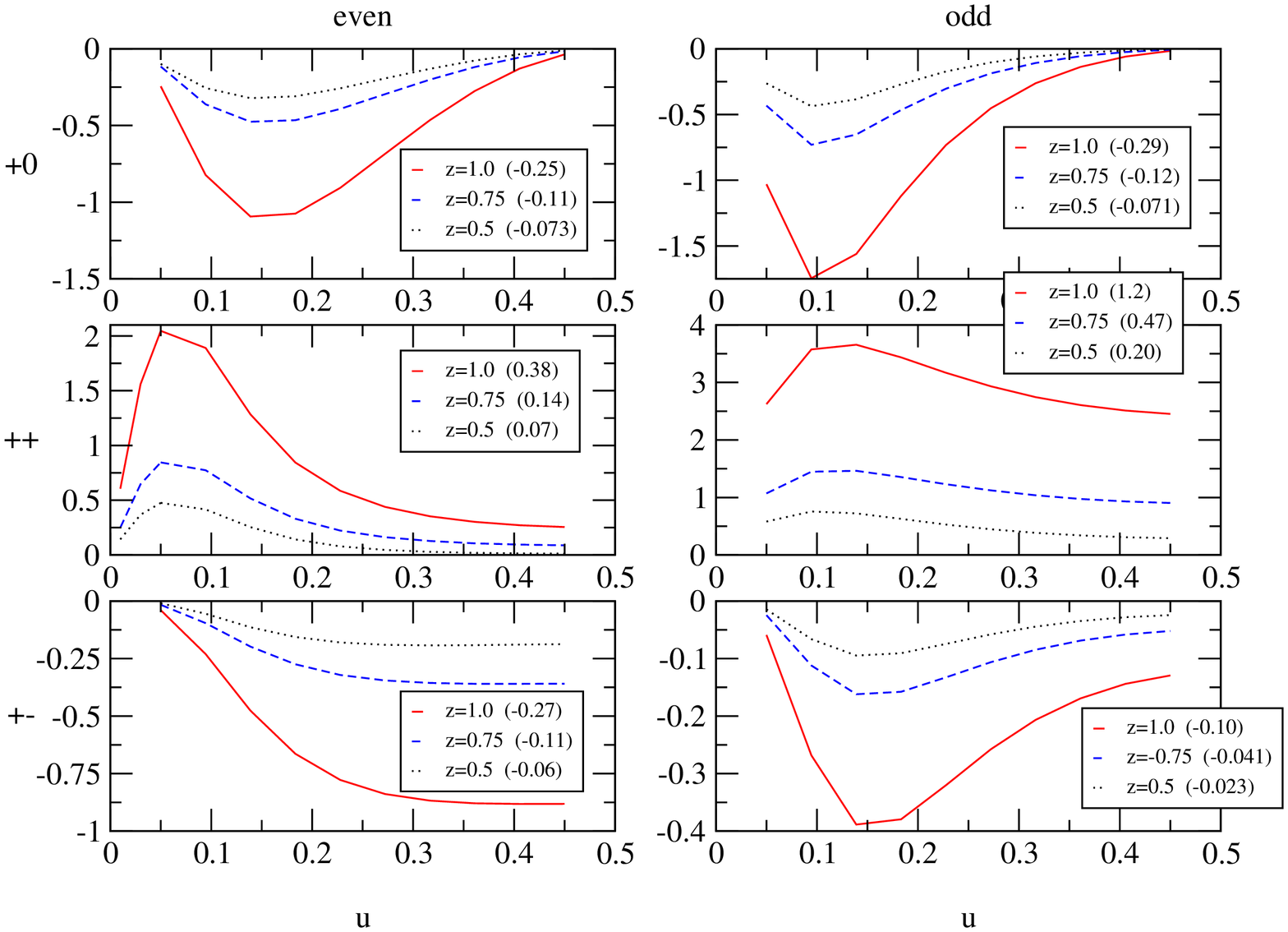}} \par}
\caption{The sensitivity to varying $z$: the six helicity amplitudes differential in $u$. The $y$-axis is $\frac{1}{C_\rho}\frac{d{M}_{+i}^{odd/even}}{du}$, where $C_{\rho}=i\hat{s}\alpha_s^{IF^2} (N^2-1/N^2)eQ_\rho f_\rho$ and $(m,\,|{\bf q}|)=(m_\rho/2,\,\sqrt{10}\,\text{GeV})$.}
\label{mz}

\end{figure}

Note that the rapidity variable $z$ 
increases (or decreases) as we raise $\alpha_s^{BFKL}$ (or $\gamma$). It is 
really this $z$ variable that we have been investigating in the
previous subsections. Figure \ref{mz} 
contains a plot for each helicity amplitude at the rapidity values, 
$z=0.5,\,0.75$ and $1.0$. The number in the brackets in the legends 
correspond to the areas underneath the curves.  Note that the position 
of the peaks do not shift as we change rapidity. As $z$ decreases all 
amplitudes fall, but a look at the integrated values demonstrates that the 
longitudinal amplitudes fall more slowly. If we lower $z$ we raise the 
ratio of the longitudinal to the transverse contributions.

\begin{figure}

{\par\centering {\includegraphics[height=10.5cm,angle=-90]{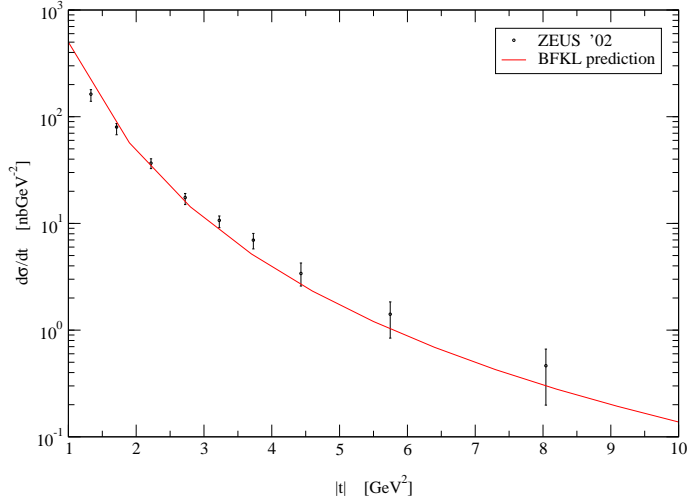}} \par}
\caption{An improved fit: $d\sigma/dt$ for parameter values, 
 $(\alpha_s^{IF},\,\alpha_s^{BFKL},\,\Lambda^2,\,m) = (0.17,\,0.25,\,m_\rho^2-t,\,m_\rho/2)$}
\label{stplay}

\end{figure}

\begin{figure}

{\par\centering {\includegraphics[height=10.5cm,angle=-90]{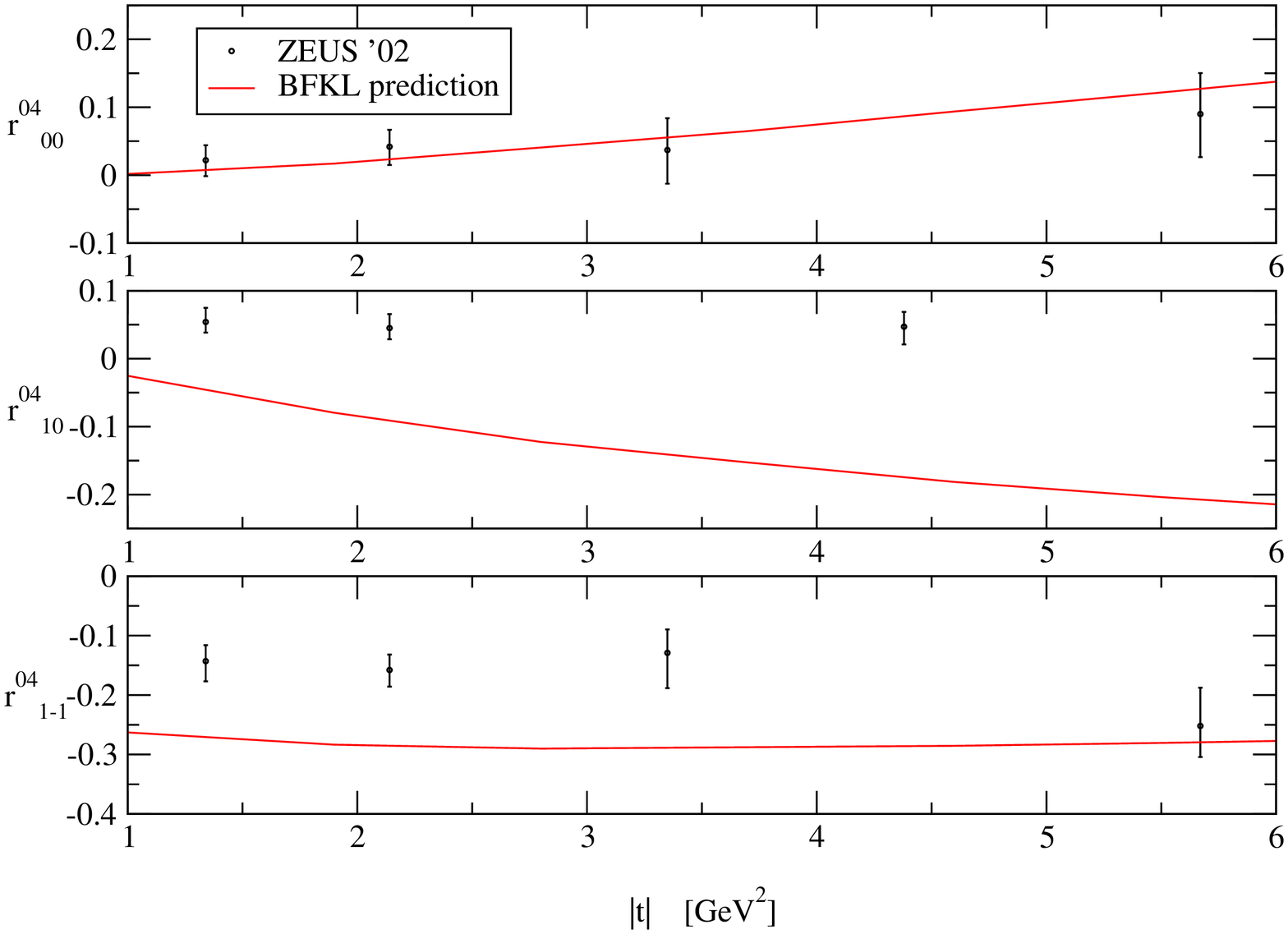}} \par}
\caption{An improved fit: $r$-matrix elements for parameter values, 
 $(\alpha_s^{BFKL},\,\Lambda^2,\,m) = (0.25,\,m_\rho^2-t,\,m_\rho/2)$.}
\label{sdmplay}

\end{figure}

This observation allows us to play off the effects of varying 
$\alpha_s^{BFKL}$ and $\Lambda^2$ simultaneously. A non-zero 
$\gamma$ value in $\Lambda^2$ improves the $t$ dependence of the 
differential cross-section and in conjunction, increasing $\alpha_s^{BFKL}$ 
drives the longitudinal component down. The fit values  
$(\alpha_s^{IF},\,\alpha_s^{BFKL},\,\Lambda^2)$ $=$ 
$(0.17,\,0.25,\,m_\rho^2-t)$ therefore lead to an improved overall fit. 
The predictions for these values are shown in Figures 
\ref{stplay}--\ref{sdmplay}.
Let us here recall the familiar result that the dependence of all amplitudes 
on the partonic collision energy $\hat s$ is characterised by
the BFKL exponent, i.e. $M_{ij} \sim \hat s^ \lambda$
with $\lambda = 12 \ln 2\, \alpha_s^{BFKL} / \pi$, provided that
$\hat s$ is sufficiently large, say $z>1$.

%%%%%%%%%%%%%%%%%%%%%%%%%%%%%%%%%%%%%%%%%%%%%%%%%%%%%%%%%%%%


%%%%%%%%%%%%%%%%%%%%%%%%%%%%%%%%%%%%%%%%%%%%%%%%%%%%%%%%%%%

\subsubsection{Sensitivity to quark mass}

\begin{figure}

{\par\centering {\includegraphics[height=10.5cm,angle=-90]{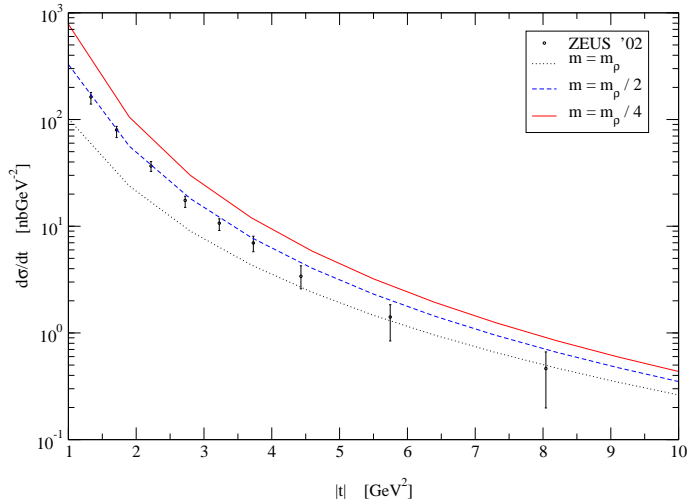}} \par}
\caption{The sensitivity to varying $m$: $d\sigma/dt$ for parameter values, 
 $(\alpha_s^{IF},\,\alpha_s^{BFKL},\Lambda^2)$ $=$ $ (0.17,\,0.2,\,m_\rho^2)$. }
\label{stmass}

\end{figure}

The sensitivity of our BFKL predictions to the quark mass parameter has 
not so far been explored. We do so now. 
Figure \ref{stmass} shows differential
 cross-section predictions. The central curve is that of the default set, 
where $(\alpha_s^{IF},\,\alpha_s^{BFKL},\,\Lambda^2,\,m)$ $=$ 
$(0.17,\,0.2,\,m_\rho^2,\,m_\rho/2)$. The other two curves were 
generated by doubling and halving the constituent quark mass. Bearing in 
mind the fact that we have varied the mass parameter considerably, we see 
that the cross-section has a fairly robust prediction, its shape being 
fairly insensitive to changes in $m$.

\begin{figure}

{\par\centering {\includegraphics[height=10.5cm,angle=-90]{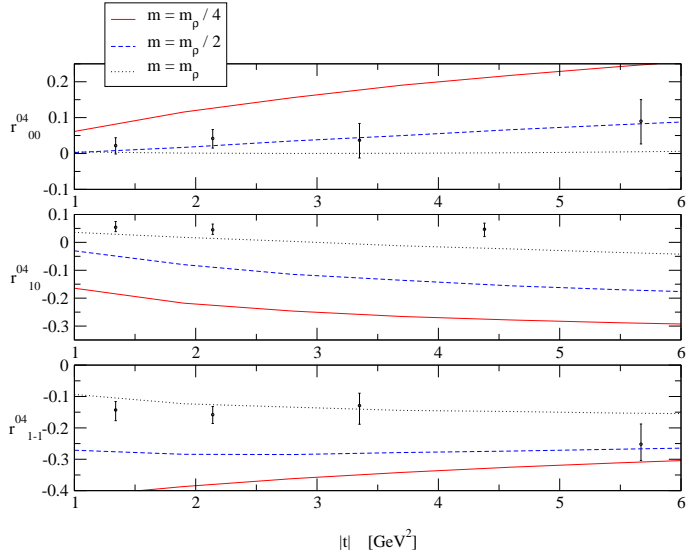}} \par}
\caption{The sensitivity to varying $m$: $r$-matrix elements for parameter values, 
 $(\alpha_s^{BFKL},\Lambda^2)$ $ =$ $(0.2,\,m_\rho^2)$.}
\label{sdmmass}

\end{figure}

\begin{figure}

{\par\centering {\includegraphics[height=10.5cm,angle=-90]{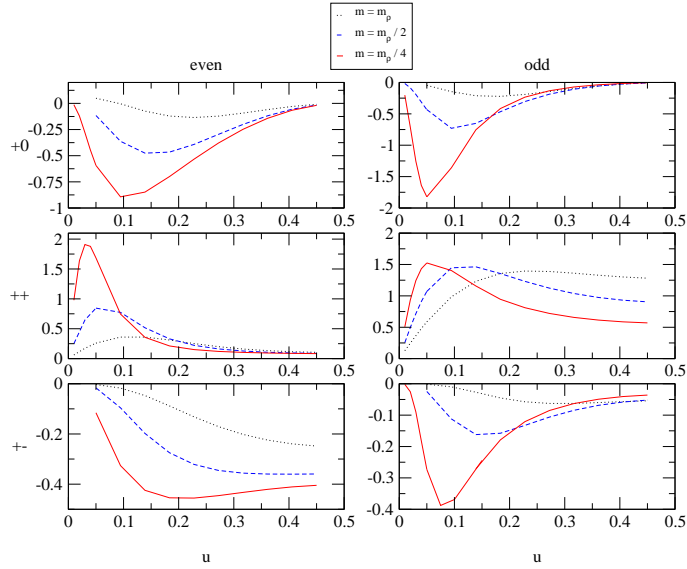}} \par}
\caption{The sensitivity to varying $m$: the six helicity amplitudes differential in $u$. The $y$-axis is $\frac{1}{C_\rho}\frac{d{M}_{+i}^{odd/even}}{du}$, where $C_{\rho}=i\hat{s}\alpha_s^{IF^2} (N^2-1/N^2)eQ_\rho f_\rho$ and $(z,\,|{\bf q}|)=(0.75,\,\sqrt{10}\,\text{GeV})$.}
\label{mm}

\end{figure}

Figure \ref{sdmmass} shows the $r$-matrix element predictions for
various quark masses. They are all sensitive to this parameter, both 
in shape and in normalisation. 
Figure \ref{mm} helps to explain why this is so. It shows the six 
helicity amplitudes, differential in $u$, for 
$m=m_\rho/4,\,m_\rho/2$ and $m_\rho$. We point out that it is not in fact the 
value of $m$ that is significant, it is the value of $m/|{\bf q}|$, and in 
these plots we fixed $|{\bf q}|=\sqrt{10}$ GeV. Notice that as we increase 
the scale $m/|{\bf q}|$, the peaks of the distributions become concentrated 
increasingly toward $u=1/2$, as expected. However, the effect 
is quite dramatic. 
Halving the constituent mass strongly enhances the contributions close to 
the end-points. The effect is particularly marked for the longitudinal 
helicity amplitudes, since they are forced to zero at exactly $u=1/2$. 
Raising the quark mass is therefore a mechanism for suppressing the 
longitudinal components with respect to the transverse. 
The dramatic dependence of the end-point region on the quark
mass, especially for the $M_{++}^{even}$, $M_{+0}^{odd}$ and $M_{+-}^{odd}$
amplitudes, is closely related to the observations of \cite{IKSS} who
found that these three amplitudes are divergent in the two-gluon
exchange approximation due to divergent end-point behaviour. We here
note that although our amplitudes are no longer divergent, even in the
massless quark limit, they do retain a memory of these large end-point
contributions. We explore the end-point region in much more detail in
Appendix B. In passing we note that this physics is reminiscent of
that for the DIS process 
$\gamma e\rightarrow \gamma \gamma^* e\rightarrow ef\bar{f}$, where $f$ is 
a fermion. The structure function $F_A^\gamma$ is zero at $u=1/2$ and has 
large contributions close to the end-points that become suppressed as we 
increase the fermion mass \cite{NS}. 

It is interesting to note that for a large 
enough $m$, the ${M}_{+0}$ amplitude not only is small but also 
changes sign. Then, the prediction of $r_{10}^{04}$ flips sign to that of 
the data. Figure \ref{sdmmass} shows this begins to occur for $m$ as low 
as $m_\rho$, in the low $|t|$ region. The evidence of this can also be seen 
even at large $t$, in Figure \ref{mm}, where the values of ${M}_{+0}$ 
differential in $u$, become positive for small $u$. So, increasing $m$ would 
be a possible mechanism for improving our fits, however, we find it 
difficult to justify a mass parameter so far away from the constituent mass. 

Increasing the mass has the effect of suppressing large dipole 
configurations due to the McDonald Bessel function, $K_{i}(m|{\bf r}|)$, 
present in the ${\bf r}$ space representation of the photon wavefunction
(e.g. see (\ref{ialbe})) which has a large argument suppression 
$\sim e^{-m|{\bf r}|}$. Figure \ref{mm} illustrates the point; 
larger masses suppress the end-point contributions more strongly. 
End-point contributions in $u$ correspond to large dipole sizes and the data 
may therefore be interpreted as suggesting that large dipole sizes 
should be more heavily suppressed than our calculations predict. 
The $\delta$-function distribution amplitude that we used in the collinear 
approximation is an extreme way of achieving this suppression 
(by forcing $u=1/2$). We suggest that the real physics of this suppression
arises as a result of Sudakov factors associated with the quark lines
\cite{BS}. 


\subsection{Sensitivity to distribution amplitudes}

In the next part of our analysis, we examine the effect of subasymptotic
corrections to the meson distribution amplitudes. We use the full
distribution amplitudes of \cite{BBKT,BB} presented in Appendix \ref{DistA} 
in conjunction with the helicity amplitudes (\ref{+0ev})--(\ref{+-odd}).
We remind the reader that we do not adjust any of the parameters of the 
distribution amplitudes. 

\begin{figure}
{\par\centering {\includegraphics[height=10.5cm,angle=-90]{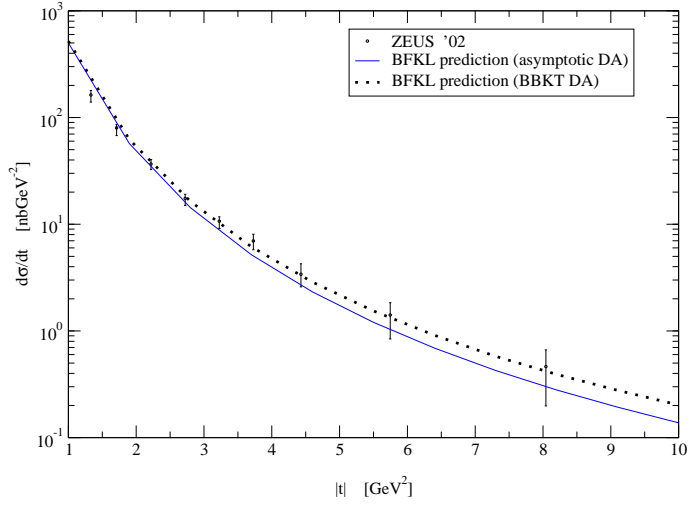}} \par}
\caption{The effect sub-asymptotic distribution amplitudes \cite{BBKT,BB}: 
$d\sigma/dt$ for parameter values, 
 $(\alpha_s^{IF},\,\alpha_s^{BFKL},\,\Lambda^2,\,m)$ $ =$ $ (0.17,\,0.25,\, m_\rho^2-t,\,m_\rho/2)$}
\label{stBB}
\end{figure}

\begin{figure}
{\par\centering {\includegraphics[height=10.5cm,angle=-90]{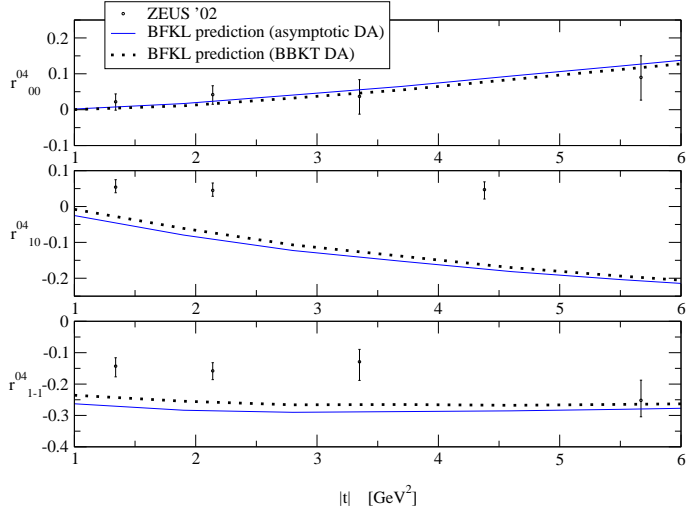}} \par}
\caption{The effect of sub-asymptotic distribution amplitudes \cite{BBKT,BB}: $r$-matrix elements for parameter values, 
 $(\alpha_s^{BFKL},\,\Lambda^2,\,m)$ 
 $=$  
$(0.25,\, m_\rho^2-t,\,m_\rho/2)$}
\label{sdmBB}
\end{figure}

Figure \ref{stBB} shows the differential cross-section in two 
distribution amplitude prescriptions which we label ``asymptotic'' and
``BBKT'' (Ball, Braun, Koike, Tanaka). The more accurate distribution 
amplitudes prove to make no qualitative difference. 
Figure \ref{sdmBB}, for the $r$ matrix elements, show that the effects 
are quite modest in these ratios. 

\begin{figure}
{\par\centering {\includegraphics[height=10.5cm,angle=-90]{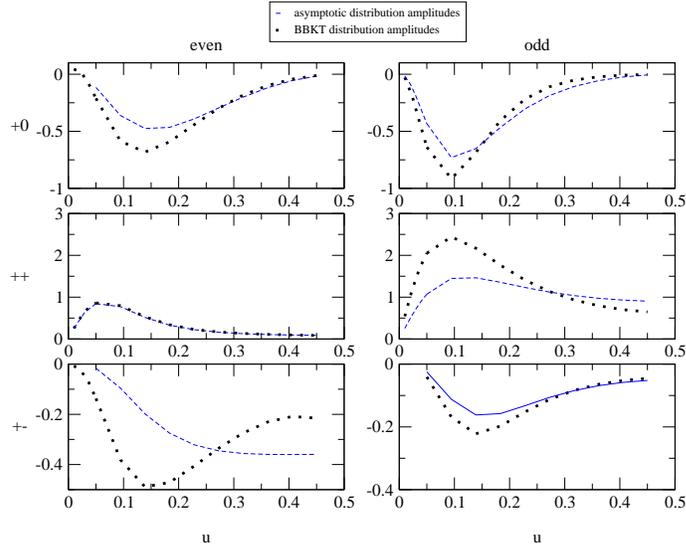}} \par}
\caption{The effect of sub-asymptotic distribution amplitudes \cite{BBKT,BB}: 
the six helicity amplitudes differential in $u$. The $y$-axis is 
$\frac{1}{C_\rho}\frac{d{M}_{+i}^{odd/even}}{du}$, where 
 $C_{\rho}=i\hat{s}\alpha_s^{IF^2} (N^2-1/N^2)eQ_\rho f_\rho$ 
 and $(z,\,m,\,|{\bf q}|)=(0.75,\,m_\rho/2,\,\sqrt{10}\,\text{GeV})$.}
\label{mBB}
\end{figure}

Figure \ref{mBB} shows the six helicity amplitudes, differential in $u$, 
at fixed rapidity $z=0.75$. Referring back to the amplitudes 
${M}_{++}^{even}$ and ${M}_{+-}^{even}$, given by (\ref{++ev}) and 
(\ref{+-ev}), we see that the difference between their $u$ dependence is a 
relative switch in sign between two pieces (one corresponding to the vector 
component of the Fierz decomposition and the other to the pseudovector). 
The plots in Figure \ref{mBB} 
demonstrate a cancellation in the effects of the non-asymptotic distribution 
amplitudes for ${M}_{++}^{even}$, but an amplification for ${M}_{+-}^{even}$. 
The effects of the different distribution amplitudes on the $u$ distributions 
of the amplitudes are most significant for ${M}_{+-}^{even}$ and 
${M}_{++}^{odd}$. However, even then there is a significant cancellation 
in their integrated values.  
We note that subasymptotic corrections are no aid to improving our fit
to $r_{10}^{04}$.


Before proceeding, we ought to comment that 
in our approach it is assumed that the dipoles which
scatter elastically at high $t$ are small in comparison to the meson
size. Therefore the dependence of the vector meson wavefunction
on the dipole size is neglected. On the other hand, we use the
perturbative photon wavefunction, with a constituent quark mass
giving the upper cut-off of the dipole sizes in the photon
(recall, that $K_a(mr) \to \exp(-mr)$ for $mr \gg 1$).
Sensitivity of the amplitudes to this parameter turns out
to be significant. This suggests that an additional suppression
of larger dipoles by the transverse part of the vector meson wave
function could well have a sizeable effect.

\section{Phenomenology for the $\rho$, $\phi$ and $J/\psi$}

We now turn our attention to the $\phi$ and $J/\psi$, reverting to the use 
of the asymptotic distribution amplitudes since wish to treat the three 
mesons on the same footing. Given the smallness of the sub-asymptotic 
corrections for the $\rho$ (as illustrated in the previous section) this
seems quite reasonable.

To what extent do the observations, made for the $\rho$ in the preceding 
sections, hold for the other mesons? The calculation for a general 
meson, $V$, with asymptotic distribution amplitude, requires only the 
substitution $(f_\rho/f_\rho^T,\,Q_\rho,\,m_\rho)$ 
$\rightarrow$ $(f_V/f_V^T,\,Q_V,\,m_V)$. The values for the meson decay
constant ($f_V$) and electromagnetic coupling ($Q_V$) are constants of 
proportionality. $m_V$ appears in the BFKL logarithm and the $\gamma V$ impact 
factor (through the quark mass $m=m_V/2$), and is qualitatively the most 
significant parameter. The mass of the $\phi$ is similar to that of the 
$\rho$, while the mass of the $J/\psi$ is significantly larger. We might 
therefore expect the predictions for the $\phi$ to be, qualitatively, similar 
to the $\rho$, and those for the $J/\psi$ to be, perhaps, very different. 
This is in fact what we shall see.

\begin{figure}
{\par\centering {\includegraphics[height=10.5cm,angle=-90]{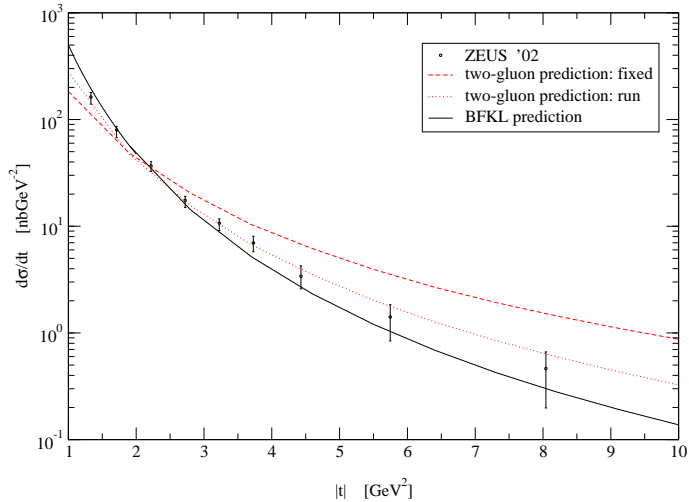}} \par}
\caption{$\rho$ photoproduction: $d\sigma/dt$.  The fixed two-gluon curve was calculated with the parameter values  $(\alpha_s^{IF},\,f_\rho,\,f_\rho^T,\,m)$
 $=$ $(0.27,\,0.216\,\text{GeV},\,$ $0.160\,\text{GeV},\,m_\rho/2)$.
 The run two-gluon curve was calculated with the parameter values 
 $(\alpha_s^{IF}(1\,\text{GeV}),\,f_\rho,\,f_\rho^T,\,m)$
=$(0.30,\,0.216\,\text{GeV},\,0.160\,\text{GeV},\,m_\rho/2)$.
 The BFKL calculated for the parameter values
 $(\alpha_s^{IF},\,\alpha_s^{BFKL},\,\Lambda^2,\,f_\rho,\,f_\rho^T,\,m)$
 $=$ $ (0.17,\,0.25,\,m_\rho^2-t,\, 0.216\,\text{GeV},\, $ $0.160\,\text{GeV},\,m_\rho/2)$.}
\label{strho}
\end{figure}

\begin{figure}
{\par\centering {\includegraphics[height=10.5cm,angle=-90]{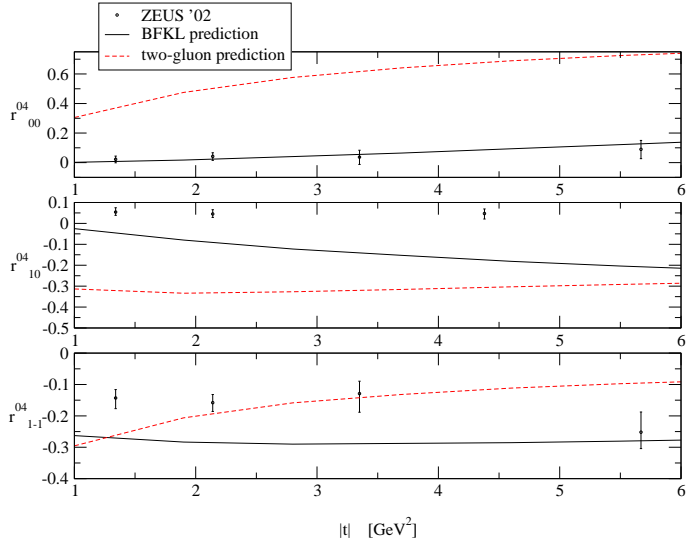}} \par}
\caption{$\rho$ photoproduction: $r$-matrix elements. The two-gluon curve was calculated with the parameter values  $(f_\rho,\,f_\rho^T,\,m)$
=
$(0.216\,\text{GeV},\,0.160\,\text{GeV},\,m_\rho/2)$. The BFKL curve was calculated for the parameter values
 $(\alpha_s^{BFKL},\,\Lambda^2,\,f_\rho,\,f_\rho^T,\,m) $
 $=$ 
$ (0.25,\,m_\rho^2-t,\, 0.216\,\text{GeV},\, $ $0.160\,\text{GeV},\,m_\rho/2)$.}
\label{sdmrho}
\end{figure}

\begin{figure}
{\par\centering {\includegraphics[height=8cm,angle=90,
viewport=216 433 450 755,clip]{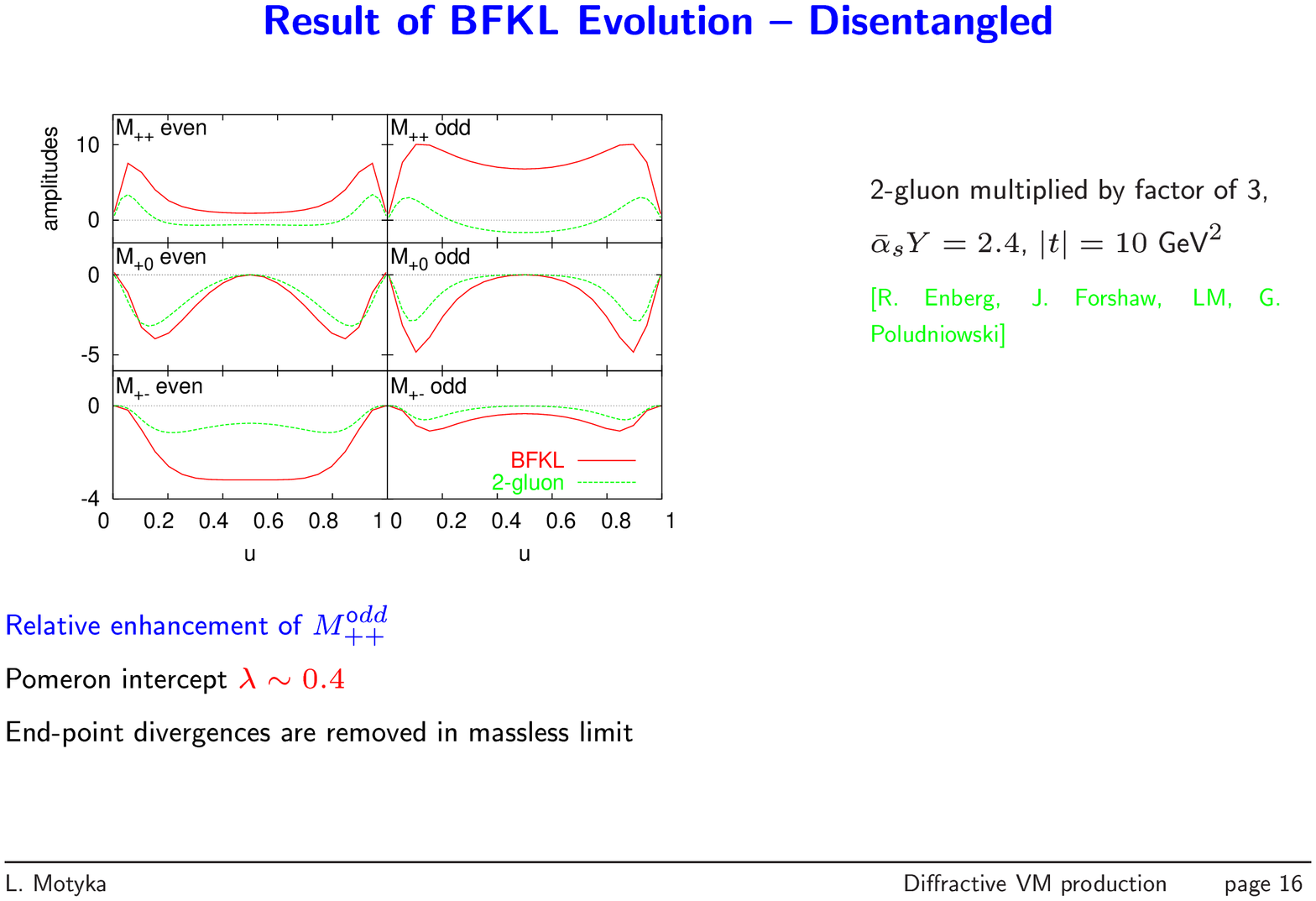}} 
\par}
\caption{The relative contributions for two-gluon exchange 
and full BFKL ($z=1.2$, \newline $t=-10$~GeV$^2$):
the six helicity amplitudes differential in $u$. Note that the two-gluon 
exchange results have been multiplied by a factor of 3.
}
\label{twogluez}
\end{figure}

Figures \ref{strho} and \ref{sdmrho} reproduce our `improved fit' 
prediction for the $\rho$ shown in Figure \ref{stplay}, 
where we played off the effects or adjusting 
$\alpha_s^{BFKL}$ and $\Lambda^2$ to our advantage. Now we also show the 
two-gluon exchange predictions. Two-gluon exchange with a fixed strong 
coupling predicts a differential cross-section far too flat in $|t|$. 
Running the coupling solves this problem and provides a good fit. 
However, looking at the two-gluon predictions for the 
$r$-matrix elements (which are independent of how we treat $\alpha_s^{IF}$), 
we see that the values for the $r_{00}^{04}$ far exceed those constrained 
by the data. This means that the longitudinal component dominates 
for two-gluon exchange. This is in line with our observation that the 
longitudinal fraction increases as we lower the $z$ rapidity (recall that 
the BFKL solution tends to two-gluon exchange in the limit $z\rightarrow 0$)
and is illustrated explicitly in Figure \ref{twogluez}. 
Thus, even though we have succeeded in getting the two-gluon curve to 
agree with the data for $d\sigma/dt$ it is generated by fundamentally 
wrong dynamics. We do however note the improvement with respect to working 
within the collinear and $\delta$-function distribution amplitude
approximation. In that prescription, two-gluon exchange was 
studied in \cite{FP,FR}. The dip to zero manifest at 
$|t|=m_V^2$ is not seen in data and is a somehwat artificial prediction 
arising from the crudity of the treatment of the meson distribution amplitudes.

\begin{figure}
{\par\centering {\includegraphics[height=10.5cm,angle=-90]{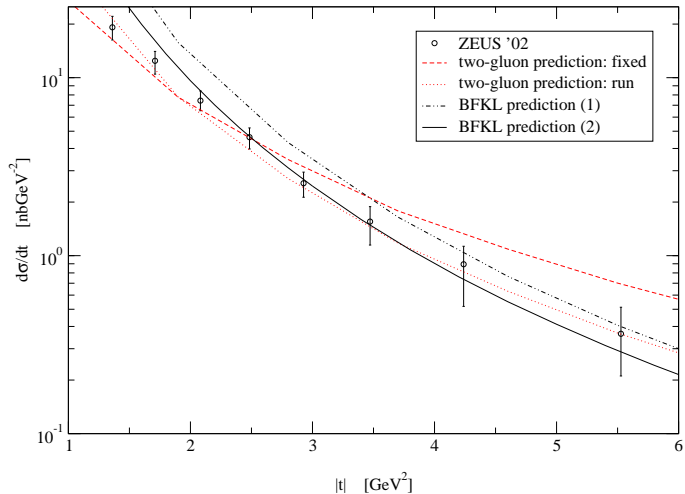}} \par}
\caption{$\phi$ photoproduction: $d\sigma/dt$ for parameter values. 
The fixed two-gluon curve was calculated with the parameter values  
 $(\alpha_s^{IF},\,f_\phi,\,f_\phi^T,\,m)$
 $=$
$(0.25,\,0.231\,\text{GeV},\,$ $0.215\,\text{GeV},\,m_\phi/2)$.
The run two-gluon curve was calculated with the parameter values  
 $(\alpha_s^{IF}(1\,\text{GeV}),\,$ $f_\phi,\,f_\phi^T,\,m)$
 $=$
 $(0.28,\,0.231\,\text{GeV},\,$ $0.215\,\text{GeV},\,m_\phi/2)$.
 BFKL prediction $(1)$ is for the parameter values
 $(\alpha_s^{IF},\,\alpha_s^{BFKL},\,$ $\Lambda^2,\,f_\phi,\,f_\phi^T,\,m)$
 $= $
$(0.17,\,0.25,\,m_\phi^2-t,\, $ $0.231\,\text{GeV},\, $ $0.215\,\text{GeV},\,m_\phi/2)$.  BFKL prediction $(2)$ is for the parameter values
 $(\alpha_s^{IF},\,\alpha_s^{BFKL},\,\Lambda^2,\,f_\phi,\,f_\phi^T,\,m)$
 $=$
 $(0.17,\,0.25,\,m_\phi^2-t,\,$ $ 0.231\,\text{GeV},\, $ $0.171\,\text{GeV},\,m_\phi/2)$.}
\label{stphi}
\end{figure}

\begin{figure}
{\par\centering {\includegraphics[height=10.5cm,angle=-90]{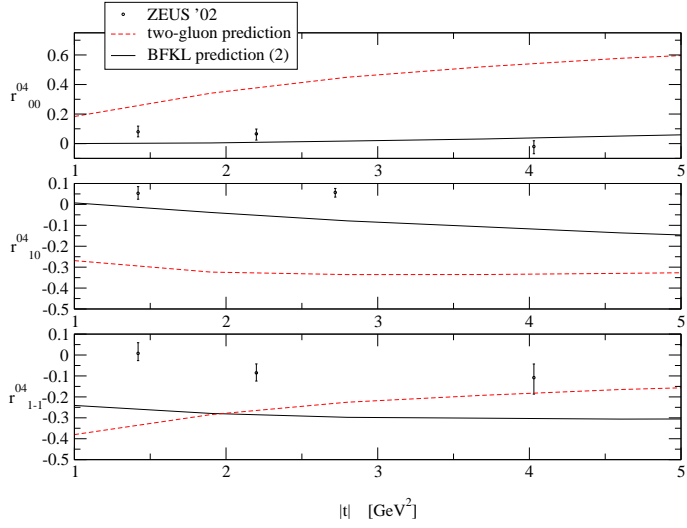}} \par}
\caption{$\phi$ photoproduction: $r$-matrix element.  The fixed two-gluon curve was calculated with the parameter values $(f_\phi,\,f_\phi^T,\,m)$
 $=$
$(0.231\,\text{GeV},\,$ $0.215\,\text{GeV},\,m_\phi/2)$.
 The BFKL prediction ($(2)$) is for the parameter values
 $(\alpha_s^{BFKL},\,\Lambda^2,\,$ $f_\phi,\,f_\phi^T,\,m)$
 $=$
$ (0.25,\,m_\phi^2-t,\, $ $0.231\,\text{GeV},\,$ $ 0.171\,\text{GeV},\,m_\phi/2)$.}
\label{sdmphi}
\end{figure}

\begin{figure}
{\par\centering {\includegraphics[height=10.5cm,angle=-90]{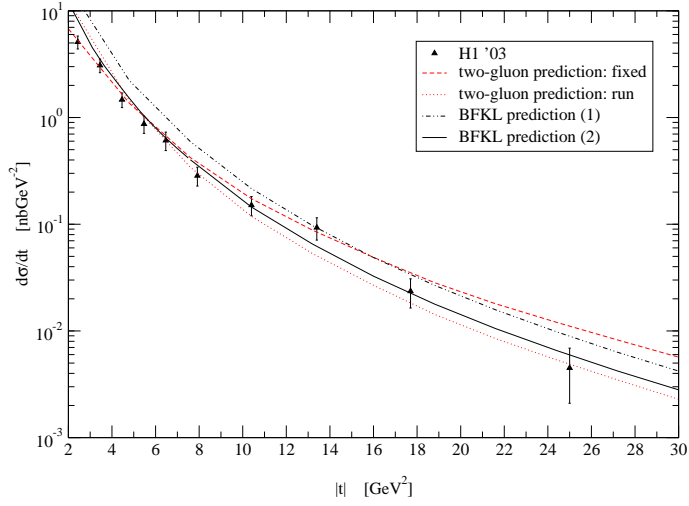}} \par}
\caption{$J/\psi$ photoproduction: $d\sigma/dt$. The fixed two-gluon curve was calculated with the parameter values 
$(\alpha_s^{IF},\,f_{J/\psi},\,f_{J/\psi}^T,\,m)$
 $=$
$(0.23,\,0.405\,\text{GeV},\,$ $0.405\,\text{GeV},\,m_{J/\psi}/2)$.
 The run two-gluon curve was calculated with the parameter values $(\alpha_s^{IF}(1\,\text{GeV})\,$ $f_{J/\psi},\,f_{J/\psi}^T,\,m)$
 $=$
$(0.29,\,0.405\,\text{GeV},\,$ $0.405\,\text{GeV},\,m_{J/\psi}/2)$.
  BFKL prediction $(1)$ is for the parameter values
 $(\alpha_s^{IF},\,\alpha_s^{BFKL},\,$ $\Lambda^2,\,$ $f_{J/\psi},\,f_{J/\psi}^T,\,m) $
 $=$
$ (0.17,\,0.25,\,m_{J/\psi}^2-t,\, 0.405\,\text{GeV},\, $ $0.405\,\text{GeV},\,m_{J/\psi}/2)$. 
 BFKL prediction $(2)$ is for the parameter values
 $(\alpha_s^{IF},\,\alpha_s^{BFKL},\,\Lambda^2,\,$ $f_{J/\psi},\,f_{J/\psi}^T,\,m) $
 $=$
$ (0.17,\,0.25,\,m_{J/\psi}^2-t,\, 0.405\,\text{GeV},\, $ $0.300\,\text{GeV},\,m_{J/\psi}/2)$.}
\label{stpsi}
\end{figure}

\begin{figure}
{\par\centering {\includegraphics[height=10.5cm,angle=-90]{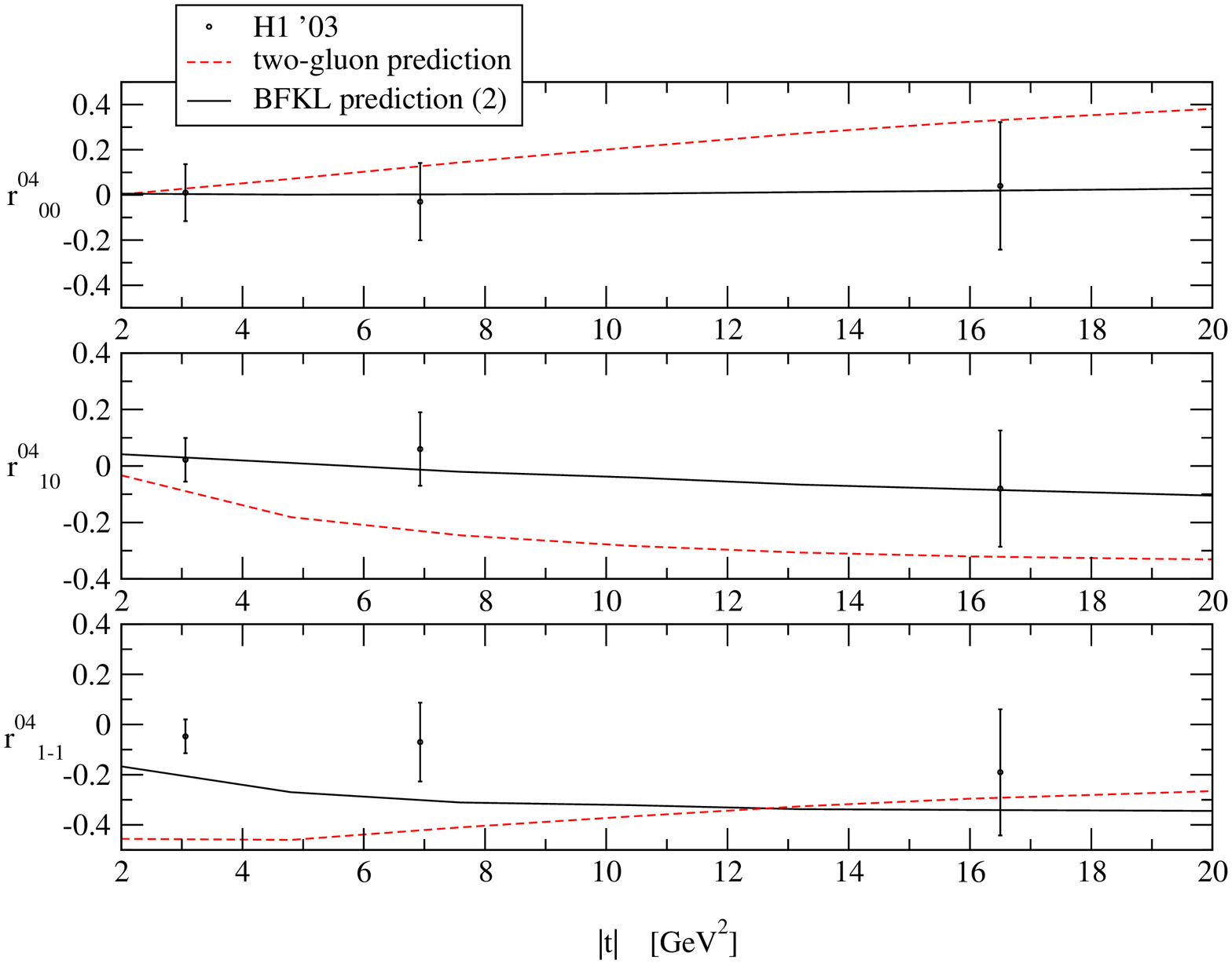}} \par}
\caption{$J/\psi$ photoproduction: $r$-matrix elements. The two-gluon curve was calculated with the parameter values 
$(f_{J/\psi},\,f_{J/\psi}^T,\,m)$
 $=$
$(0.405\,\text{GeV},\,$ $0.405\,\text{GeV},\,m_{J/\psi}/2)$.
 The BFKL prediction ($(2)$) is for the parameter values
 $(\alpha_s^{BFKL},\,\Lambda^2,\,$ $f_{J/\psi},\,f_{J/\psi}^T,\,m)$
 $=$
$ (0.25,\,m_{J/\psi}^2-t,\, 0.405\,\text{GeV},\, $ $0.300\,\text{GeV},\,m_{J/\psi}/2)$. }
\label{sdmpsi}
\end{figure}

Figures \ref{stphi}--\ref{sdmphi} and \ref{stpsi}--\ref{sdmpsi} 
present our results for the $\phi$ and $J/\psi$ mesons. 
The BFKL curves which we label ``(1)'' in Figures \ref{stphi} and 
\ref{stpsi} were obtained by simply making the appropriate change in meson 
constants and masses. No further adjustment of the other parameters is made
relative to the fit for the $\rho$ meson.
For the curves labelled ``BFKL (2)'' we 
altered the values of the meson tensor coupling but have kept the ratio 
\begin{eqnarray}
f_V^T=\frac{f_\rho^T}{f_\rho}f_V.      
\end{eqnarray}
Fixing $f_V^T$ in this way improves the fit to data. Note that if we 
allowed ourselves slightly more freedom with the values of $f_V^T$, the 
quality of the fits to the $t$-distribution could be improved still further.

The $\phi$ 
predictions are similar to those for the $\rho$. We again see that the 
two-gluon exchange $t$-distribution requires a running strong coupling to 
fit the data and that the $r$-matrix elements point to this being
an accident. 
The BFKL predictions are again correctly dominated by the transverse 
contributions, but again fail to predict the correct sign for $r^{04}_{10}$ 
in the case of the $\phi$. However, for the $J/\psi$, the BFKL predictions 
are compatible with all the observables. The large quark mass drives the 
longitudinal amplitudes small enough to agree with the $r^{04}_{10}$ data. 
We note that although the two-gluon exchange $r$-matrix predictions for 
the $J/\psi$ are in marginal agreement with the data, we have to doubt the 
validity of the underlying dynamics, given the comments above.

Comparing two-gluon exchange and BFKL predictions, we see that the BFKL 
calculation is a definite step forward. 
The success of the our predictions for the $\phi$ and $J/\psi$ is a 
noteworthy result. We now understand the success of the apparently naive
analysis of \cite{FP} as being due to the fact that it correctly 
identified the dominance of the 
${M}^{odd}_{++}$ amplitude. 
However, the inability of BFKL to agree with the data for $r_{10}^{04}$ 
suggests that while the longitudinal contribution is brought down by BFKL 
effects, it is still not under enough control to accurately describe all 
the observables measured at ZEUS and H1. 




\section{Summary}

We have compared our theoretical predictions for $d\sigma/dt$ and the three 
$r$-matrix elements to the data of ZEUS and H1, in various levels of 
approximation. The only meson we found complete agreement for, in any scheme, 
was the $J/\psi$. We nevertheless could generally obtain good fits to the 
$r_{00}^{04}$ observable. Fits to the cross-sections favoured running the 
BFKL scale in a natural way: $\Lambda^2=m_V^2-t$. The theoretical  
predictions for $r^{04}_{10}$ were the main obstacles to obtaining 
satisfactory fits for the $\rho$ and $\phi$. We found the natural 
predictions of both two-gluon and BFKL pomeron exchange to be the wrong sign.
The predictions of the observables proved insensitive to higher twist terms, 
suggesting that these corrections are unlikely to provide a solution. We 
also found that using the distribution amplitudes of \cite{BBKT,BB} had 
little effect on the quality of fits compared to the asymptotic ones. The 
inability of theoretical predictions to fit $r_{10}^{04}$, and the general 
sensitivity of the $r$-matrix elements to the quark mass parameter, $m$, 
therefore seem to locate the problem at the level of the hard subprocess. 
We suggest that the data hint that we may require a greater suppression of 
larger dipoles than we predict and that this may be provided by Sudakov 
suppression of emissions off the quark lines.

We note the dominance of the ${M}_{++}^{odd}$ contribution 
and have demonstrated that this corresponds to the amplitude used in the 
BFKL calculation of \cite{FP} (which was, in addition, conducted in a 
$\delta$-function approximation). We were able to get excellent agreement 
for the $\phi$ and $J/\psi$ differential cross-sections 
$using$ $the$ $same$ $parameters$ that fitted the $\rho$. On the whole the 
evidence suggests that leading logarithm BFKL predictions work well for the 
transverse amplitudes but, despite suppressing the longitudinal component 
relative to Born level, it fails to have enough control over this 
contribution.

\section*{Dedication and Acknowledgements}
We should like to dedicate this paper to the memory of Jan Kwiecinski,
a good friend, colleague and inspirational teacher.

RE wishes to thank the Theoretical Physics Group at the University of 
Manchester for their hospitality when parts of this work was carried out.
LM is grateful to Gunnar Ingelman and the Uppsala THEP group for their warm
hospitality.
This research was funded in part by the UK Particle Physics and
Astronomy Research Council (PPARC), by the Swedish Research Council, and
by the Polish Committee for Scientific Research (KBN) grant no.\ 5P03B~14420.

\appendix
\section{Vector meson distribution amplitudes} \label{DistA}

The authors of \cite{BBKT} presented results for the 
complete set of distribution amplitudes up to twist-3 for the $\rho$ and 
$\phi$ (and $K$) \cite{BBKT}. These results were extended to twist-4 in
\cite{BB}. The relevant distribution amplitudes, up to and including twist-4, 
are classified in Table \ref{DA}.
\begin{table}
\begin{center}
\begin{tabular}{|c||c|c|c|} \hline
 & Twist 2 & Twist 3 & Twist 4  \\
 & $O(1)$ & $O(1/q)$ & $O(1/q^2)$ \\ \hline
$\|$ & $\phi_\|$ & $h_\|^{(t)},\,h_\|^{(s)}$ &  \\ 
$\perp$ & $\phi_\perp$ & $g_\perp^{(v)},\,g_\perp^{(a)}$ & $h_3$ \\ \hline
\end{tabular}
\end{center}
\caption{Classification of distribution amplitudes up to twist-4.}
\label{DA}
\end{table}

We quote all parameters at the renormalisation scale 
$\mu^2=1\,\text{GeV}^2$ and neglect the slow running.
In the $q\rightarrow \infty$ limit, the parameters 
vanish. We refer to this as the asymptotic limit, and the distribution 
amplitudes in this limit as the asymptotic distribution amplitudes. 

Tables \ref{DAE} and \ref{DAO} hold the values of all the relevant 
parameters, at the scale $\mu^2=1\,\text{GeV}^2$.
\begin{table}
\begin{center}
\begin{tabular}{|c|c|c|c|c|c|} \hline \hline
 & $a_1^\|$ & $a_2^\|$ & $\zeta_3$ & $\omega_3^A$ & $\omega_3^V$ \\ \hline 
$\rho$ & $0$ & $0.18\pm 0.10$ & $0.032\pm 0.010$ & $-2.1\pm 1.0$ & $3.8\pm 
1.8$ \\ \hline
\end{tabular}
\end{center}
\caption{Chiral even parameters: $\mu^2=1\,\text{GeV}^2$.}
\label{DAE}
\end{table}
\begin{table}
\begin{center}
\begin{tabular}{|c|c|c|c|c|c|c|} \hline \hline
 & $a_1^\perp$ & $a_2^\perp$ & $\zeta_3$ & $\omega_3^T$ & $\zeta_4^T$ & $\tilde{\zeta}_4^T$ \\ \hline 
$\rho$ & $0$ & $0.20\pm 0.10$ & $0.032\pm 0.010$ & $7.0\pm 7.0$ 
& $0.10\pm 0.05$ 
& $-0.10\pm 0.05$ \\ \hline
\end{tabular}
\end{center}
\caption{Chiral odd parameters: $\mu^2=1\,\text{GeV}^2$.}
\label{DAO}
\end{table}
We do not include quark mass corrections in this study. They are in any
case zero for the $\rho$ meson.

\subsection{Twist-2 distribution amplitudes}

\subsubsection{Chiral Even}

There is one chiral even, twist-2 amplitude:
\begin{eqnarray}
\phi_\|(u)=6u\bar{u}\left(1+3a_1^\|\xi+a_2^\|\,\frac{3}{2}(5\xi^2-1) \right).
\end{eqnarray}
and the asymptotic distribution is
\begin{eqnarray}
\phi_\|^{asy}(u)=6u\bar{u}.
\end{eqnarray}

\subsubsection{Chiral Odd}

There is one chiral odd, twist-2 amplitude. This can be written
\begin{eqnarray}
\phi_\perp(u)=6u\bar{u}\left(1+3a_1^\perp\xi+a_2^\perp\,\frac{3}{2}(5\xi^2-1) \right).
\end{eqnarray}
The asymptotic distribution is
\begin{eqnarray}
\phi_\|^{asy}(u)=6u\bar{u}.
\end{eqnarray}

\subsection{Twist-3 distribution amplitudes}

\subsubsection{Chiral Even}

There are two even, twist-3 distributions:
\begin{eqnarray}
g_\perp^{(a)}(u)=6u\bar{u}\left[1+\left\{\frac{1}{4}a_2^\|+\frac{5}{3}\zeta_3\left(1-\frac{3}{16}\omega_3^A+\frac{9}{16}\omega_3^V\right)\right\}(5\xi^2-1) \right]
\end{eqnarray}
and
\begin{eqnarray}
&& g_\perp^{(v)}(u)=\frac{3}{4}(1+\xi^2)+(\frac{3}{7}a_2^\|+5\zeta_3)\,(3\xi^2-1)\nonumber \\ && +\left[\frac{9}{112}a_2^\|+\frac{15}{64}\zeta_3(3\omega_3^V-\omega_3^A) \right](3-30\xi^2+35\xi^4)
\end{eqnarray}
where the subscript of `$3$' refers to three-particle corrections and we 
neglect mass corrections. In the asymptotic limit
\begin{eqnarray}
g_\perp^{(a)\,asy}(u)=6u\bar{u}.
\end{eqnarray}
and
\begin{eqnarray}
g_\perp^{(v)\,asy}(u)=\frac{3}{4}\,(1+\xi^2).
\end{eqnarray}

\subsubsection{Chiral Odd}
There is one relevant chiral odd twist-3 distribution:
\begin{eqnarray}
h_\|^{(t)}(u)=3\xi^2+\frac{3}{2}\,a_2^\perp\,\xi^2(5\xi^2-3)+\frac{15}{16}\,\zeta_3\,\omega_3^T(3-30\xi^2+35\xi^4).
\end{eqnarray}
In the asymptotic limit
\begin{eqnarray}
h_\|^{(t)\,asy}(u)=3\xi^2.
\end{eqnarray}

\subsection{Twist-4 distribution amplitudes}

\subsubsection{Chiral Odd}

There is one relevant chiral odd twist-4 distribution;
\begin{eqnarray}
&& h_3(u)=1+\left(-1+\frac{3}{7}a_2^\perp-10(\zeta_4^T+\tilde{\zeta}_4^T)\right)C_2^{1/2}(\xi)\nonumber \\ && \hspace{4cm}
+\left(-\frac{3}{7}a_2^\perp-\frac{15}{8}\zeta_3\omega_3^T\right)C_4^{1/2}(\xi).
\end{eqnarray}
where $C_n^\lambda(\xi)$ are Gegenbauer polynomials.
In the asymptotic limit
\begin{eqnarray}
&& h_3^{asy}=1-C_2^{1/2}(\xi)=\frac{3}{2}(1-\xi^2).
\end{eqnarray}

\subsection{Distribution amplitude schemes}

We have six helicity amplitudes to consider, i.e. 
(\ref{+0ev})--(\ref{+-odd}). Their individual dependence on 
the distribution amplitudes is as follows:
\begin{eqnarray}
\Phi_{+0}^{even}(u)\equiv\phi_\|(u)
\end{eqnarray}
\begin{eqnarray}
\Phi_{++}^{even}(u)\equiv\left(\frac{g_\perp^{(a)}(u)}{4}-(1-2u)\int_0^udv\,(\phi_\|(v)-g_\perp^{(v)}(v))\right)
\end{eqnarray}
\begin{eqnarray}
\Phi_{+-}^{even}(u)\equiv\left(\frac{g_\perp^{(a)}(u)}{4}+(1-2u)\int_0^udv\,(\phi_\|(v)-g_\perp^{(v)}(v))\right)
\end{eqnarray}
\begin{eqnarray}
\Phi_{+0}^{odd}(u)\equiv\int_0^udv\,\left(h_\|^{(t)}(v)-\phi_\perp(v) \right)
\end{eqnarray}
\begin{eqnarray}
\Phi_{++}^{odd}(u)\equiv\phi_\perp(u)
\end{eqnarray}
\begin{eqnarray}
\Phi_{+-}^{odd}(u)\equiv\int_0^udv\int_0^vd\eta\left(h_\|^{(t)}(\eta)-\frac{1}{2}\phi_\perp(\eta)-\frac{1}{2}h_3(\eta) \right).
\end{eqnarray}
We now present the explicit formulae for these amplitudes in four different 
prescriptions. 

\subsubsection{Prescription 1: leading twist (collinear) approximation 
with $\delta$-function distributions}
We keep only the twist-2 contributions and equate the corresponding 
distribution amplitudes to a $\delta$-function that enforces the quark and 
antiquark share the meson momentum equally. Then
\begin{eqnarray}
\Phi_{+0}^{even}(u)=\delta\left(u-\frac{1}{2}\right)
\end{eqnarray}
\begin{eqnarray}
\Phi_{++}^{even}(u)=0
\end{eqnarray}
\begin{eqnarray}
\Phi_{+-}^{even}(u)=0
\end{eqnarray}
\begin{eqnarray}
\Phi_{+0}^{odd}(u)=0
\end{eqnarray}
\begin{eqnarray}
\Phi_{++}^{odd}(u)=\delta\left(u-\frac{1}{2}\right)
\end{eqnarray}
\begin{eqnarray}
\Phi_{+-}^{odd}(u)=0.
\end{eqnarray}

\subsubsection{Prescription 2: leading twist (collinear) approximation with 
asymptotic distribution amplitudes}
We keep only the twist-2 contributions and equate the corresponding 
distribution amplitudes to their asymptotic forms. Then
\begin{eqnarray}
\Phi_{+0}^{even}(u)=6u\bar{u}
\end{eqnarray}
\begin{eqnarray}
\Phi_{++}^{even}(u)=0
\end{eqnarray}
\begin{eqnarray}
\Phi_{+-}^{even}(u)=0
\end{eqnarray}
\begin{eqnarray}
\Phi_{+0}^{odd}(u)=0
\end{eqnarray}
\begin{eqnarray}
\Phi_{++}^{odd}(u)=6u\bar{u}
\end{eqnarray}
\begin{eqnarray}
\Phi_{+-}^{odd}(u)=0.
\end{eqnarray}

\subsubsection{Prescription 3: higher twist approximation with asymptotic 
distribution amplitudes}
We keep all twist and equate the 
corresponding distribution amplitudes to the asymptotic forms. After 
integration:
\begin{eqnarray}
\Phi_{+0}^{even}(u)=6u\bar{u}
\end{eqnarray}
\begin{eqnarray}
&& \Phi_{++}^{even}(u)=
3u\bar{u}(u^2+\bar{u}^2)
\end{eqnarray}
\begin{eqnarray}
&& \Phi_{+-}^{even}(u)=
6u^2\bar{u}^2
\end{eqnarray}
\begin{eqnarray}
&& \Phi_{+0}^{odd}(u)=3(1-2u)u\bar{u}
\end{eqnarray}
\begin{eqnarray}
&& \Phi_{+-}^{odd}(u)=\frac{3}{2}u^2\bar{u}^2
\end{eqnarray}

\subsubsection{Prescription 4: higher twist approximation with BBKT 
distribution amplitudes}
We keep all twist and equate the corresponding 
distribution amplitudes to the BBKT forms listed in A1--A3.
We neglect to explicitly quote the trivial, but bulky, integrated forms.

\section{End-point behaviour} \label{ppendpt}
\subsection{Two-gluon exchange and Reggeisation corrections}
The importance of contributions at, and close to, end-points is a matter of 
debate (see for example (\cite{IKSS,Hoyer,IK})). 
The authors of \cite{IKSS} worked in the two-gluon exchange
approximation in the light-quark limit\footnote{We refer to our previous
paper \cite{EFMP} for a discussion on this limit.}
and concluded that $M_{++}^{even}$, $M_{+0}^{odd}$ and $M_{+-}^{odd}$ 
were all plagued by end-point divergences. They argued that these
end-point divergences would be brought under control by Sudakov corrections
and hence that the dominant amplitude would remain $M_{++}^{odd}$. 
In \cite{Hoyer}, the end-point divergences were interpreted as the signal
that QCD factorisation was breaking down.
In the context of a specific model of the vector 
meson wavefunction, \cite{IK} obtain a meson wavefunction,
dependent on both longitudinal and transverse coordinates,
by a Borel transform of the photon wavefunction.
%In their analysis, configurations with large invariant masses of the 
%quark--anti-quark
%pair get exponentially suppressed. This feature seems to be
%plausible from the point of view of the local parton hadron duality
%hypothesis and it connects also to the Martin-Ryskin-Teubner
%model of the meson wavefunction \cite{MRT}.
The model of \cite{IK} gives an explicit form for the end-point 
contributions at all
twist, it is shown that those contributions can be resummed
and that the breakdown of factorisation by end-point divergencies
at leading twist is only apparent. Here we discuss the role of end-points
within the context of leading logarithmic BFKL resummation.

\begin{figure}
{\par\centering {\includegraphics[height=10cm,angle=-90]{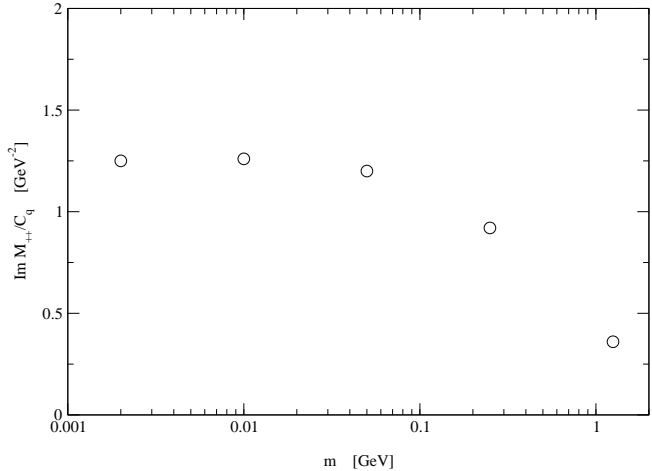}} \par}
\caption{The ${M}_{++}^{even}$ amplitude against $m$, the quark mass 
parameter. 
The fixed parameter values are $(z,m_V,|t|)=(1.0,m_\rho,10\, \text{GeV}^2)$. 
The constant, $C_q=s\alpha_s^2(N^2-1/N^2)f_Vm_Vee_q$ and the asymptotic 
distribution amplitudes were used.}
\label{mass}
\end{figure}

In Figure \ref{stmass}, 
we also observe the appearance of large end-point 
contributions (in the same amplitudes noted by \cite{IKSS}). 
It is the case that 
these contributions lead to divergent matrix elements in the two-gluon 
approximation in the case of massless quarks. However, these divergences
are not present after re-summing the BFKL logarithms. In Figure \ref{mass}
we show that the 
the BFKL prediction for the $(++)$ amplitude remains finite even in the 
massless quark limit. We shall investigate how this divergence in (++) at 
the Born level arises and how higher orders bring it under control. The 
authors of \cite{IKSS} worked with the asymptotic distribution amplitudes; 
we shall also do so (see Appendix \ref{DistA} for the asymptotic formulae). 
The end-point behaviour is independent of this approximation. 

The light-quark limit corresponds to putting the mass of the quarks that 
form the meson to zero. This leads to the replacements
\begin{eqnarray}
&& mK_1(m|r|)\rightarrow 1/|r|\hspace{0.5cm}\text{and} \hspace{0.5cm}
K_0(m|r|)\rightarrow -m\text{ln}(m|r|)\sim 0,
\end{eqnarray}
where we have assumed that $m|r|\rightarrow 0$, i.e. that we get 
no significant contributions from asymptotically large transverse sizes. 

We can apply the light quark approximation to the Born (two-gluon) level 
amplitudes previously derived. For $M^{even\,2g}_{++}$ one has
\begin{eqnarray}
&& {M}^{even\,2g}_{++} =
-sC_Vf_VM_V \frac{6\pi}{|{\bf q}|^4}
\left( \text{ln}\left(\frac{1-u_{min}}{u_{min}}\right)-1+2u_{min}\right).
\label{umindiv}
\end{eqnarray}
A clear divergence as the cut-off $u_{min} \to 0$.

The full BFKL solution has proven in the past to improve 
infra-red finiteness of scattering amplitudes; see for example 
\cite{MT,MMR} 
for the case of $qq\rightarrow qq$ scattering. At any 
finite order in $\alpha_s$ 
this scattering amplitude diverges. Even in infra-red finite observables 
the Born level result can be seen to give undue weight to 
contributions from momentum configurations which are unphysical,
i.e. where all the momentum is short-circuited down one gluon. 

The authors of \cite{FR} demonstrated how this problem can be resolved 
without recourse to the full BFKL result, but rather by a fixed order 
resummation of a subset of the BFKL logarithms. Following \cite{FR}, we
go from the Born amplitude to the resummed one by the substitution
\begin{eqnarray}
\frac{1}{{\bf {\bf k}}^2({\bf {\bf k}}-{\bf {\bf q}})^2}\rightarrow \frac{1}{{\bf {\bf k}}^2({\bf {\bf k}}-{\bf {\bf q}})^2}\left( \frac{{\bf {\bf k}}^2({\bf {\bf k}}-{\bf {\bf q}})^2}{{\bf q}^4} \right)^z.
\end{eqnarray}
The substitution above clearly suppresses small gluon momenta and the
formerly divergent amplitude can now be written 
(using the definition of the incomplete beta function)
\begin{eqnarray}
&& {M}^{even\,2gR}_{++} 
=-\frac{6\pi}{|{\bf q}|^4}sC_Vf_VM_V \nonumber \\ && \times\biggl[ B(1-u_{min},2z,1+2z)- B(1-u_{min},1+2z,1+2z)\biggr]_{u_{min}}^{1-u_{min}}.
\end{eqnarray}
We can safely put $u_{min}\rightarrow 0$ for $z> 0$: 
\begin{eqnarray}
&& {M}^{even\,2gR}_{++} =
-\frac{6\pi}{|{\bf q}|^4}sC_Vf_VM_V\,\left(B(2z,1+2z)- B(1+2z,1+2z)\right) 
\nonumber \\ &&
=-\frac{6\pi}{|{\bf q}|^4}sC_Vf_VM_V\,
\frac{\Gamma(2z)\Gamma(2+2z)}{\Gamma(2+4z)}.
\end{eqnarray}
As it should be, we recover the divergence in the limit 
$z\rightarrow 0$:\footnote{Of course we trivially reproduce the divergence
of (\ref{umindiv}) if we first take the limit $z \to 0$.}
\begin{eqnarray}
&& {M}^{even\,2gR}_{++}(z\rightarrow 0)
\rightarrow -\frac{3\pi}{|{\bf q}|^4}sC_Vf_VM_V\frac{1}{z}.
\end{eqnarray}
We observe that this is equal to that derived for $qq\rightarrow qq$ 
scattering amplitude in \cite{MT,MMR}, 
excepting a factor of $-3C_Vf_VM_V/2q^2$. This is understood since
in this limit the quark and antiquark are far apart and the dominant
contribution comes from coupling the exchanged reggeised gluons to
a single parton \cite{FR,BFLLR}. 
The clear result is that the resummed $(++)$ even amplitude 
remains finite even in the massless quark limit for non-zero rapidity. 
We have demonstrated analytically that higher orders accomplish this 
and so it is no longer surprising that 
Figure \ref{mass} demonstrates convergence at $m \to 0$.

\subsection{BFKL}
The previous subsection is sufficient to illustrate that the BFKL result
is free from infrared divergences. Here we take a closer look at the
end-point behaviour of the full BFKL amplitude. The reader not interested
in the details may wish to jump directly to the final result, (\ref{Mppu0}).
Equation (\ref{+0ev}) and the definition of the asymptotic 
distributions in Appendix \ref{DistA} give us the result
\begin{eqnarray}
&& {M}_{++}^{even}=\frac{sC_V f_V M_V}{8|q|}\,
\int_0 ^1 du\;
3\, u \bar{u}\,(u^2+\bar{u}^2)
%u\bar{u}\biggl( \int_{0}^{u}\frac{dv}{\bar{v}}\phi_{\parallel}(v)+
%\int_{u}^{1}\frac{dv}{v}\phi_{\parallel}(v)\biggr) 
\nonumber\\
&& \times \sum_{n=-\infty}^{n=+\infty}
\int_{-\infty}^{\infty}d\nu
\frac{\nu^{2}+n^{2}}{[\nu^{2}+(n-1/2)^{2}][\nu^{2}+(n+1/2)^{2}]}
\frac{\exp [\chi_{2n}(\nu)z]}{\sin (i\pi\nu)} \,
I_{00}(\nu,2n,q, u;1) \nonumber \\ \label{ppbfkl}
\end{eqnarray}
where
\begin{eqnarray}
&& I_{\alpha\beta}(\nu,n,q,u;a) =
\frac{m}{2}\int^{C^{\prime}+i\infty}_{C^{\prime}-i\infty}
\frac{d\zeta}{2\pi i}\Gamma(a/2-\zeta)\Gamma(-a/2-\zeta)\,
\tau_q ^{\zeta} \; (i\, \text{sign}\,(1-2u))^{\alpha-\beta+n}  \nonumber \\ &&
\times  \left(\frac{4}{|q|}\right)^{4}
\left[\sin\pi(\alpha + \mu + \zeta)\; \right.
B(\alpha,\mu, q^* ,u,\zeta)\,
B(\beta,\widetilde\mu,q,u^* ,\zeta) \nonumber \\ &&
\hspace{2cm} - (-1)^n
\sin\pi(\alpha - \mu + \zeta)\;
B(\alpha,-\mu, q^* ,u,\zeta)\,
B(\beta,-\widetilde\mu,q,u^* ,\zeta)
\left. \right]
\label{fintegral}
\end{eqnarray}
and
\begin{eqnarray}
B(\alpha,\mu, q^* ,u,\zeta) =  
(-4u \bar u)^{-(\mu+2+\alpha+\zeta)/2}
\left(\frac{4}{ q^* }\right)^\alpha
2^{-\mu}\,
\frac{\Gamma(\mu+2+\alpha+\zeta)}{\Gamma(\mu+1)} \nonumber \\
{}_2F_1\left(\frac{\mu+2+\alpha+\zeta}{2} \, , \,
\frac{\mu-1-\alpha-\zeta}{2}\, ; \,
\mu+1\, ; \,\frac{1}{4u \bar u}\right),
\label{blocks21}
\end{eqnarray}
with  $\tau_q = 4m^2/|q|^2$.

The even amplitudes have $a=1$ so the integrand in $I_{\alpha\beta}(\nu,n,q,u;1)$ has a pole at $\zeta=-1/2$, where $-1<C^\prime<-1/1$. If we shift the contour to the right we pick up pole contributions from $\text{Re}\,\zeta\ge-1/2$. In the limit $m\rightarrow 0$ we might expect the leading pole to be that at $\zeta=-1/2$, the residue of which is independent of $m$, the other poles contributions vanishing in the limit. The integrand is complicated, however, expressed how it is in terms of blocks of hypergeometric functions. It is difficult to see clearly the analytic behaviour in $\zeta$. For example the true expansion parameter may not be $4m^2/q^2$, rather something like $4m^2/u^2q^2$, where it is not clear whether the `leading pole' actually is leading in the simultaneous limit $m\rightarrow 0$ and $u\rightarrow 0$. We shall see if some manipulation of the formula can be enlightening.
Using the following hypergeometric identity
\begin{eqnarray}
 {}_2F_1(a,b;c;z)=&& \frac{\Gamma(c)\Gamma(b-a)}{\Gamma(b)\Gamma(c-a)}(-z)^{-a}\,{}_2F_1(a,a+1-c;a+1-b;1/z) \nonumber \\ &&
+\frac{\Gamma(c)\Gamma(a-b)}{\Gamma(a)\Gamma(c-b)}(-z)^{-b}\,{}_2F_1(b,b+1-c;b+1-a;1/z),
\end{eqnarray} 
we can re-write,
\begin{eqnarray}
&& B(\alpha,\mu, q^* ,u,\zeta) =  
(-4u \bar u)^{-(\mu+2+\alpha+\zeta)/2}
\left(\frac{4}{ q^* }\right)^\alpha
2^{-\mu}\,
\Gamma(\mu+2+\alpha+\zeta) \nonumber \\ && \times
\biggl[ \frac{\Gamma(-3/2-\alpha-\zeta)}{\Gamma\left(\frac{\mu-1-\alpha-\zeta}{2}\right)\Gamma\left( \frac{\mu-\alpha-\zeta}{2} \right)}\left( -\frac{1}{4u\bar{u}}\right)^{-\frac{\mu+2+\alpha+\zeta}{2}}\nonumber \\ && \times
 {}_2F_1\left( \frac{\mu+2+\alpha+\zeta}{2} ,\frac{2+\alpha+\zeta-\mu}{2} , 5/2+\alpha+\zeta ,4u\bar{u}\right) \nonumber \\ &&
+\frac{\Gamma(3/2+\alpha+\zeta)}{\Gamma\left(\frac{\mu+2+\alpha+\zeta}{2}\right)\Gamma\left( \frac{\mu+3+\alpha+\zeta}{2} \right)}\left( -\frac{1}{4u\bar{u}}\right)^{-\frac{\mu-1-\alpha-\zeta}{2}}\nonumber \\ && \times
{}_2F_1\left( \frac{\mu-1-\alpha-\zeta}{2} ,-\frac{1+\alpha+\zeta+\mu}{2} , -1/2-\alpha-\zeta ,4u\bar{u}\right) \biggr].
\end{eqnarray}
The advantage of this expression is that the hypergeometrics now have well defined power series expansions for $0<u<1$. We can simplify this expression further using the gamma function identity,
\begin{eqnarray}
\frac{1}{\Gamma(z)\Gamma(z+1/2)}=\frac{2^{2z-1}}{\sqrt{\pi}}\frac{1}{\Gamma(2z)}.
\end{eqnarray}
We find,
\begin{eqnarray}
&&  B(\alpha,\mu, q^* ,u,\zeta) = \left(\frac{2}{q^*}\right)^\alpha\frac{1}{\sqrt{\pi}}2^{-2-\zeta} \nonumber \\ && \times\left[
C(\mu,\alpha,|u|,\zeta)+(-u\bar{u})^{-3/2-\alpha-\zeta}D(\mu,\alpha,|u|,\zeta),
 \right]
\end{eqnarray}
where we define
\begin{eqnarray}
&& C(\mu,\alpha,|u|,\zeta)=\frac{\Gamma(\mu+2+\alpha+\zeta)\Gamma(-3/2-\alpha-\zeta)}{\Gamma(\mu-1-\alpha-\zeta)}\nonumber \\ && \times {}_2F_1\left( \frac{\mu+2+\alpha+\zeta}{2} ,\frac{2+\alpha+\zeta-\mu}{2} , 5/2+\alpha+\zeta ,4u\bar{u}\right)
\end{eqnarray}
and
\begin{eqnarray}
&&D(\mu,\alpha,|u|,\zeta)=\Gamma(3/2+\alpha+\zeta) \nonumber \\ && 
\hspace{1cm}\times 
{}_2F_1\left( \frac{\mu-1-\alpha-\zeta}{2} ,-\frac{1+\alpha+\zeta+\mu}{2} , -1/2-\alpha-\zeta ,4u\bar{u}\right).
\end{eqnarray}
Note that both the $C$ and $D$ functions are single-valued in $u$, but that the $B$ function explicitly breaks single-valuedness. This is not a problem, since it is only the combination of blocks that must be single-valued. The blocks occur in the combination
\begin{eqnarray}
&& \text{sin}\,\pi(\alpha+\mu+\zeta)\,B(\alpha,\mu,q^*,u\zeta)B(\beta,\tilde{\mu},q,u^*,\zeta) \nonumber \\ &&
\hspace{2cm}-\,(-1)^n\text{sin}\,\pi(\alpha-\mu+\zeta)\,B(\alpha,-\mu,q^*,u\zeta)B(\beta,-\tilde{\mu},q,u^*,\zeta) \nonumber \\ &&
=\text{sin}\,\pi((\alpha+\mu+\zeta))\,\left(\frac{2}{q^*}\right)^\alpha \left(\frac{2}{q}\right)^\beta \frac{1}{\pi} 2^{-4-2\zeta} \nonumber \\ && \times\biggl[C(\mu,\alpha,|u|,\zeta)C(\tilde{\mu},\beta,|u|,\zeta) \nonumber \\ && \hspace{1cm}+C(\mu,\alpha,|u|,\zeta)D(\tilde{\mu},\beta,|u|,\zeta)\,(u\bar{u})^{-3/2-\beta-\zeta}e^{\pm\pi(3/2+\beta+\zeta)}  \nonumber \\ && \hspace{1cm}+D(\mu,\alpha,|u|,\zeta)C(\tilde{\mu},\beta,|u|,\zeta)\,(u\bar{u})^{-3/2-\alpha-\zeta}e^{\mp\pi(3/2+\alpha+\zeta)}
  \nonumber \\ && \hspace{2cm}+ D(\mu,\alpha,|u|,\zeta)D(\tilde{\mu},\beta,|u|,\zeta)\,(u\bar{u})^{-3-\alpha-\beta-2\zeta}(-1)^{\alpha-\beta}\biggr] \nonumber \\ &&
-\,(-1)^n\text{sin}\,\pi(\alpha-\mu+\zeta)\,\left(\frac{2}{q^*}\right)^\alpha \left(\frac{2}{q}\right)^\beta \frac{1}{\pi} 2^{-4-2\zeta}\,\biggl[ \mu,\,\tilde{\mu}\rightarrow -\mu,\,-\tilde{\mu}\biggr].\nonumber \\
\end{eqnarray}
This expression is complicated but in fact simplifies. Note that in the square bracket there are two multi-valued pieces corresponding to the cross multiplication of $C$'s and $D$'s. The sum of the two multi-valued pieces are not single-valued together in general; they have different $u$-dependences. In fact each multi-valued piece cancels exactly with another piece from the other square bracket\footnote{This has been verified numerically.}. We can throw away the cross multiplied terms. We obtain the result,
\begin{eqnarray}
&& I_{\alpha\beta}(\nu,n,q,u;a) =
\frac{m}{2\pi}\left(\frac{2}{q}\right)^{2+\beta}\left(\frac{2}{q^*}\right)^{2+\alpha}\left(\text{i\,sgn}\,(1-2u)\right)^{\alpha-\beta+n} 
\nonumber \\ && \hspace{2cm}
\times\int^{C^{\prime}+i\infty}_{C^{\prime}-i\infty}
\frac{d\zeta}{2\pi i}\Gamma(a/2-\zeta)\Gamma(-a/2-\zeta)\,
(m^2/|q|^2) ^{\zeta} \nonumber \\ && \hspace{-1cm}
\times \biggl[\sin\pi(\alpha + \mu + \zeta)
\,G(\alpha,\beta,\mu,\widetilde\mu,|u| ,\zeta)
\nonumber \\ && \hspace{2cm}  - (-1)^n
\sin\pi(\alpha - \mu + \zeta)
\,G(\alpha,\beta,-\mu,-\widetilde\mu,|u| ,\zeta) \biggr]
\label{fintegral2}
\end{eqnarray}
where we introduce the notation,
\begin{eqnarray}
&& G(\alpha,\beta,\mu,\widetilde\mu,|u| ,\zeta)=C(\mu,\alpha,|u|,\zeta)C(\tilde{\mu},\beta,|u|,\zeta) \nonumber \\ &&
\hspace{1.5cm}+(-1)^{\alpha-\beta}\,(u\bar{u})^{-3-\alpha-\beta-2\zeta}\,D(\mu,\alpha,|u|,\zeta)D(\tilde{\mu},\beta,|u|,\zeta).
\label{Gfor}
\end{eqnarray}
We are now in a better position to examine the behaviour of the (++) even amplitude as $u\rightarrow 0$.

Note that the full (++) even amplitude is proportional to the integral over $u$ of $I_{00}$ multiplied by the $u$-dependence arising from the distribution amplitudes. At the end-points, we get the behaviour,
\begin{eqnarray}
\frac{d{M}_{++}^{even}(u\rightarrow 0)}{du}\propto 3u\bar{u}(u^2+\bar{u}^2)\,u^{-3-2\zeta}\rightarrow 3u^{-2-2\zeta}.
\end{eqnarray}
The leading pole approximation would put $\zeta\rightarrow -0.5$, leaving the integrated amplitude logarithmically divergent as in the Born level case. However, earlier we showed that resumming leads to a finite result for non-zero rapidity. We now seek to show that the leading pole divergence, for non-zero rapidity, is an artifact indicating the break down of assumptions made rather than genuine asymptotic behaviour.

The G-function (\ref{Gfor}) simplifies greatly at the end-points. In the limit $u\rightarrow 0$ the dominant $u$-dependence allows us to write
\begin{eqnarray}
&& G(\alpha,\beta,\mu,\widetilde\mu,|u|\rightarrow 0,\zeta)\rightarrow
(-1)^{\alpha-\beta}\,u^{-3-\alpha-\beta-2\zeta}\,\Gamma(3/2+\alpha+\zeta)\Gamma(3/2+\beta+\zeta), \nonumber \\
\end{eqnarray}
where we have put the hypergeometrics to unity in this limit. This simplification leads to
\begin{eqnarray}
&& I_{\alpha\beta}^{u\rightarrow 0}(\nu,n,q,u;a) =
\frac{m}{\pi}\left(\frac{2}{q}\right)^{2+\beta}\left(\frac{2}{q^*}\right)^{2+\alpha}(-1)^{\alpha-\beta}\left(\text{i\,sgn}\,(1-2u)\right)^{\alpha-\beta+n}
\frac{\text{sin}\,\pi\mu}{u^{3+\alpha+\beta}} 
\nonumber \\ &&
\hspace{2cm}\times\int^{C^{\prime}+i\infty}_{C^{\prime}-i\infty}
\frac{d\zeta}{2\pi i}\Gamma(a/2-\zeta)\Gamma(-a/2-\zeta)\Gamma(3/2+\alpha+\zeta)\Gamma(3/2+\beta+\zeta)\nonumber \\ && \hspace{4cm}\times\text{sin}\,\pi(\alpha+1/2+\zeta)\,
(m^2/|q|^2) ^{\zeta} \nonumber \\ &&
=-m\left(\frac{2}{q}\right)^{2+\beta}\left(\frac{2}{q^*}\right)^{2+\alpha}(-1)^{\alpha-\beta}\left(\text{i\,sgn}\,(1-2u)\right)^{\alpha-\beta+n}
\frac{\text{sin}\,\pi\mu}{u^{3+\alpha+\beta}} 
\nonumber \\ &&
\hspace{3cm} \times G_{22}^{21}\left( \frac{m^2}{u^2|q|^2}|^{-1/2-\beta,\,-1/2-\alpha}_{\hspace{0.25cm}a/2,\,-a/2}\right),
\end{eqnarray}
where we have used the integral definition of the Meijer G-function 
For $(a,\alpha,\beta)=(1,0,0)$ we find that this simplifies;
\begin{eqnarray}
 G_{22}^{21}\left( \frac{m^2}{u^2|q|^2}|^{-1/2,\,-1/2}_{1/2,\,-1/2}\right)=G_{11}^{11}\left( \frac{m^2}{u^2|q|^2}|^{-1/2}_{1/2}\right)=\frac{\frac{m}{u|q|}}{\left(1+\frac{m^2}{u^2|q|^2}\right)^2}.
\end{eqnarray}
We can therefore write down the (++) even amplitude in the $u\rightarrow 0$ limit as
\begin{eqnarray}
&& \frac{dM_{++}^{even}(u\rightarrow 0)}{du}\rightarrow \frac{3C_Vf_VM_Vm^2}{8|q|^2}\left( \frac{2}{|q|}\right)^4 \frac{1}{u^3}\frac{1}{\left(1+\frac{m^2}{u^2|q|^2}\right)^2} \nonumber \\ && \times \sum_{n=-\infty}^{n=+\infty}
\int_{-\infty}^{\infty}d\nu
\frac{\nu^{2}+n^{2}}{[\nu^{2}+(n-1/2)^{2}][\nu^{2}+(n+1/2)^{2}]}
e^{\chi_{2n}(\nu)z}.
\label{Mppu0}
\end{eqnarray}
Note that the $u$-dependence has completely factorised from the $n$ and $\nu$ 
dependences. In fact the result, up to a factor, is that of 
$qq\rightarrow qq$ scattering found in \cite{MT,MMR}. 
We observed this earlier when performing the resummed calculation. 
From inspection of (\ref{Mppu0}), for any particular conformal spin, 
we can see that it vanishes as $u\rightarrow 0$ for finite $m/|q|$ and 
also for $m/|q|\rightarrow 0$ and non-zero $u$.  We know, however, that 
we must recover the Born level divergence as $z\rightarrow 0$. In fact we 
do achieve this; the sum over conformal spins is infinite for zero $z$. 
The fact that conformal spin has decoupled from $u$ in our formula seems to 
imply that for zero $z$ we face a divergence regardless of $m\rightarrow 0$. 
In fact this is not so. For sufficiently large $n$ the assumption we made in 
putting the hypergeometric functions to zero breaks down. We should have a 
cut off in $n$ relating to the inverse of $u_{min}$. 

\subsection{Conclusions}
In this appendix we have shown that
\begin{itemize}
\item the two-gluon exchange (++) even amplitude diverges as 
$\text{ln}\,u_{min}$ as $u_{min}\rightarrow 0$ for $m= 0$
\item the resummed (++) even amplitude converges as $u_{min}\rightarrow 0$ 
and $m\rightarrow 0$ for non-zero $z$, but diverges as $1/z$, 
as $z\rightarrow 0$ (the two-gluon limit).
\item the BFKL (++) even amplitude converges as $u_{min}\rightarrow 0$ 
and $m\rightarrow 0$, when $z$ is non-zero.
\item for the complete (or resummed) leading log calculation we can only 
get a divergence with the following simultaneous limits 
($u_{min},\,m,\,z)\rightarrow 0$ and any one can provide a cut-off.
\end{itemize}


\begin{thebibliography}{99}

\bibitem{EFMP}
R.~Enberg, J.~R.~Forshaw, L.~Motyka and G.~Poludniowski,
%``Vector meson photoproduction from the BFKL equation. I: Theory,''
JHEP {\bf 0309} (2003) 008
%[arXiv:hep-ph/0306232].

\bibitem{BFKL}
L.~N.~Lipatov,
\emph{Sov.\ J.\ Nucl.\ Phys.\ } {\bf 23} (1976) 338;
%%CITATION = SJNCA,23,338;%%
%
E.~A.~Kuraev, L.~N.~Lipatov and V.~S.~Fadin,
\emph{Sov.\ Phys.\ JETP }{\bf 44}, 443 (1976); {\it ibid.\ } {\bf 45} (1977) 199;
%%CITATION = SPHJA,44,443;%%
%%CITATION = SPHJA,45,199;%%
%
I.~I.~Balitsky and L.~N.~Lipatov,
\emph{Sov.\ J.\ Nucl.\ Phys.\ }{\bf 28} (1978) 822.
%%CITATION = SJNCA,28,822;%%
%[arXiv:hep-ph/0107068].
%
L.~N.~Lipatov,
Sov.\ Phys.\ JETP {\bf 63} (1986) 904;
\emph{Phys.\ Rep.\ } {\bf 286} (1997) 131.
%%CITATION = HEP-PH 9610276;%%

\bibitem{FR}
J.~R.~Forshaw and M.~G.~Ryskin,
\emph{Z.\ Phys.\ }{\bf C68} (1995) 137.
%%CITATION = HEP-PH 9501376;%%

\bibitem{ZEUS1}
S.~Chekanov {\it et al.}  [ZEUS Collaboration],
%``Measurement of proton dissociative diffractive photoproduction of  vector mesons at large momentum transfer at HERA,''
\emph{Eur.\ Phys.\ J.\ }{\bf C26} (2003) 389.
%[arXiv:hep-ex/0205081].
%%CITATION = HEP-EX 0205081;%%

\bibitem{Aktas:2003zi}
A.~Aktas {\it et al.}  [H1 Collaboration],
%``Diffractive photoproduction of J/psi mesons with large momentum  transfer at HERA,''
Phys.\ Lett.\ B {\bf 568} (2003) 205
%[arXiv:hep-ex/0306013].

\bibitem{IKSS}
D.~Y.~Ivanov, R.~Kirschner, A.~Sch\"afer and L.~Szymanowski,
%``The light vector meson photoproduction at large t,''
\emph{Phys.\ Lett.\ }{\bf B478} (2000) 101
[Erratum-\emph{ibid.\ }{\bf B498} (2001) 295].
%[arXiv:hep-ph/0001255].
%%CITATION = HEP-PH 0001255;%%

\bibitem{Hoyer}
P.~Hoyer, J.~T.~Lenaghan, K.~Tuominen and C.~Vogt,
%``Subprocess size in hard exclusive scattering,''
arXiv:hep-ph/0210124.
%%CITATION = HEP-PH 0210124;%%

\bibitem{FP}
J.~R.~Forshaw and G.~Poludniowski,
%``Vector meson photoproduction at high-t and comparison to HERA data,''
\emph{Eur.\ Phys.\ J.\ }{\bf C26} (2003) 411.

\bibitem{Critt}
J.A. Crittenden, ``Exclusive Production of Neutral Vector Mesons at the 
Electron-proton Collider HERA'', Springer Tracts in Modern Physics v.140, 
Springer-Verlag (1997).

\bibitem{SSW}
K. Schilling, P. Seyboth, G.E. Wolf, Nucl. Phys. {\bf B15} (1970) 397 [Erratum-ibid. {\bf B18} (1970) 332. 

\bibitem{SW}
K. Schilling and G.E. Wolf, Nucl.Phys. {\bf B61} (1973) 381. 

\bibitem{BBKT}
P.~Ball, V.~M.~Braun, Y.~Koike and K.~Tanaka,
%``Higher twist distribution amplitudes of vector mesons in {QCD}: Formalism  and twist three distributions,''
\emph{Nucl.\ Phys.\ }{\bf B529} (1998) 323.
%[arXiv:hep-ph/9802299].
%%CITATION = HEP-PH 9802299;%%

\bibitem{BB}
P.~Ball and V.~M.~Braun,
%``Higher twist distribution amplitudes of vector mesons in {QCD}: Twist-4 distributions and meson mass corrections,''
\emph{Nucl.\ Phys.\ }{\bf B543} (1999) 201.
%[arXiv:hep-ph/9810475].
%%CITATION = HEP-PH 9810475;%%

\bibitem{BFKLP}
S.J. Brodsky, V.S. Fadin, V.T. Kim, L.N. Lipatov and G.B. Pivovarov, 
JETP Lett.70 (1999) 155. 
 
\bibitem{CFL}
B.E. Cox, J.R. Forshaw and L. L\"onnblad, JHEP 9910:023.

\bibitem{HVM} 
R.~Enberg, L.~Motyka and G.~Poludniowski,
%``Diffractive heavy vector meson production from the BFKL equation,''
\emph{Eur.\ Phys.\ J.\ }{\bf C26} (2002) 219.
%[arXiv:hep-ph/0207027].
%%CITATION = HEP-PH 0207027;%%

\bibitem{NS}
R. Nisius and M.H. Seymour, Phys. Lett. {\bf B452} (1999) 409.

\bibitem{GPS1}
I.F.Ginzburg, S.L.Panfil and V.G.Serbo, Nuc.Phys $\mathbf{B284}$ (1987) 685.

\bibitem{BS}
J. Botts and G. Sterman, Nuc. Phys. {\bf B325} (1989) 62.

\bibitem{MT}
A.~H.~Mueller and W.-K.~Tang,
\emph{Phys.\ Lett.\ }{\bf B284} (1992) 123.
%%CITATION = PHLTA,B284,123;%%

\bibitem{MMR}
 L.~Motyka, A.~D.~Martin and M.~G.~Ryskin,
\emph{Phys.\ Lett.\ } {\bf B524} (2002) 107.
%%CITATION = HEP-PH 0110273;%%

\bibitem{BFLLR}
J.~Bartels, J.~R.~Forshaw, H.~Lotter, L.~N.~Lipatov, M.~G.~Ryskin and
M.~W\"usthoff,
\emph{Phys.\ Lett.\ }{\bf B348} (1995) 589.
%%CITATION = HEP-PH 9501204;%%

\bibitem{IK}
A.~Ivanov and R.~Kirschner,
%``Electroproduction of vector mesons: Factorization of end-point  contributions,''
Eur.\ Phys.\ J.\ C {\bf 29} (2003) 353
[arXiv:hep-ph/0301182].

%\bibitem{BFLW}
%J.~Bartels, J.~R.~Forshaw, H.~Lotter and M.~W\"usthoff,
%\emph{Phys.\ Lett.\ }{\bf B375} (1996) 301.
%%CITATION = HEP-PH 9601201;%%
%
%\bibitem{PSIPSI}
%J.~Kwieci\'{n}ski and L.~Motyka,
%%``Diffractive J/psi production in high energy gamma gamma collisions as  a probe of the {QCD} pomeron,''
%\emph{Phys.\ Lett.\ }{\bf B438} (1998) 203.
%%[arXiv:hep-ph/9806260].
%%%CITATION = HEP-PH 9806260;%%

%\bibitem{GINZ}
%I.~F.~Ginzburg, S.~L.~Panfil and V.~G.~Serbo,
%%``Possibility Of The Experimental Investigation Of The QCD Pomeron In Semihard Processes At The Gamma Gamma Collisions,''
%\emph{Nucl.\ Phys.\ }{\bf B284} (1987) 685.
%%CITATION = NUPHA,B284,685;%%

%\bibitem{GINZ2}
%I.~F.~Ginzburg, S.~L.~Panfil and V.~G.~Serbo,
%\emph{Nucl.\ Phys.\ }{\bf B296} (1987) 569.

%\bibitem{NP}
%H.~Navelet and R.~Peschanski,
%\emph{Nucl.\ Phys.\ B }{\bf 507} (1997) 353.
%%CITATION = HEP-PH 9703238;%%

%%\cite{Fadin:1998py}
%\bibitem{Fadin:1998py}
%V.~S.~Fadin and L.~N.~Lipatov,
%``BFKL pomeron in the next-to-leading approximation,''
%Phys.\ Lett.\ B {\bf 429} (1998) 127
%[arXiv:hep-ph/9802290].
%%CITATION = HEP-PH 9802290;%%

%\cite{Ciafaloni:1998gs}
%\bibitem{Ciafaloni:1998gs}
%M.~Ciafaloni and G.~Camici,
%``Energy scale(s) and next-to-leading BFKL equation,''
%Phys.\ Lett.\ B {\bf 430} (1998) 349
%[arXiv:hep-ph/9803389].
%%CITATION = HEP-PH 9803389;%%

%\cite{Enberg:2001ev}
%\bibitem{Enberg:2001ev}
%R.~Enberg, G.~Ingelman and L.~Motyka,
%``Hard colour singlet exchange and gaps between jets at the Tevatron,''
%Phys.\ Lett.\ B {\bf 524} (2002) 273
%[arXiv:hep-ph/0111090].
%%CITATION = HEP-PH 0111090;%%

%\bibitem{Brodsky1}
%S.~J.~Brodsky, V.~S.~Fadin, V.~T.~Kim, L.~N.~Lipatov and G.~B.~Pivovarov,
%``The {QCD} pomeron with optimal renormalization,''
%JETP Lett.\  {\bf 70} (1999) 155
%[arXiv:hep-ph/9901229].
%%CITATION = HEP-PH 9901229;%%

%\cite{Brodsky2}
%\bibitem{Brodsky2}
%S.~J.~Brodsky, V.~S.~Fadin, V.~T.~Kim, L.~N.~Lipatov and G.~B.~Pivovarov,
%``High-energy QCD asymptotics of photon photon collisions,''
%JETP Lett.\  {\bf 76} (2002) 249
%[Pisma Zh.\ Eksp.\ Teor.\ Fiz.\  {\bf 76} (2002) 306]
%[arXiv:hep-ph/0207297].
%%CITATION = HEP-PH 0207297;%%

%\bibitem{GR} I.~S.~Gradshteyn and I.~M.~Ryzhik, \emph{Table of Integrals, Series, and Products}, 6th Edition, Academic Press (2000).

%\bibitem{LVM}
%R.~Enberg, L.~Motyka and G.~Poludniowski,
%\emph{Acta Phys.\ Polon.\ }{\bf B33} (2002) 3511.


\end{thebibliography}
\end{document}